\documentclass[fleqn,usenatbib]{mnras}

\usepackage{newtxtext,newtxmath}

\usepackage[T1]{fontenc}
\usepackage[usenames]{color}
\usepackage{graphicx,float}
\usepackage{amsmath}	% Advanced maths commands

\usepackage{amssymb}	% Extra maths symbols
\usepackage{tabularx,ragged2e,booktabs,caption}
\usepackage{mathrsfs}
\usepackage{units}
\usepackage{booktabs,makecell}

\newcolumntype{L}{>{\raggedright\arraybackslash}X}

\usepackage{adjustbox}
\usepackage{array,booktabs}
\usepackage{orcidlink}

\usepackage[normalem]{ulem}
\newcolumntype{C}[1]{>{\centering\arraybackslash}p{#1}}

\newcommand{\zf}{$z_\mathrm{f}$}
\newcommand{\tf}{$t_\mathrm{f}$}
\newcommand{\Deltat}{$\Delta t$}
\newcommand{\Mgal}{$M_\mathrm{Gal}$}
\newcommand{\Msun}{$\mathrm{M_{\odot}}$}
\newcommand{\Z}{$Z$}
\newcommand{\Zsun}{$\mathrm{Z_{\odot}}$}

\newcommand{\zmbbh}{$z_\mathrm{mBBH}$}
\newcommand{\tmbbh}{$t_\mathrm{mBBH}$}
\newcommand{\tdelay}{$t_\mathrm{delay}$}

\newcommand{\PopAstro}{$Pop_\mathrm{Astro}$}
\newcommand{\PopBPS}{$Pop_\mathrm{BPS}$}

\newcommand{\lenPopBPS}{$n_\mathrm{BPS}$}
\newcommand{\MassPopBPS}{$M_\mathrm{BPS}$}

\newcommand{\scatter}{$\sigma$}

\begin{document}

\author[R. Srinivasan et al.]{
Rahul Srinivasan \orcidlink{0000-0002-7176-6690},$^{1,2}$ \thanks{rahul.srinivasan@oca.eu} 
Astrid Lamberts,$^{1,2}$ 
Marie Anne Bizouard
\orcidlink{0000-0002-4618-1674},$^{2}$ 
Tristan Bruel
\orcidlink{0000-0002-1789-7876},$^{1,2}$ and
\newauthor{Simone Mastrogiovanni
\orcidlink{0000-0003-1606-4183},$^{1,2,3}$}
\\
$^{1}$Universit\'e C\^ote d’Azur, Observatoire de la C\^ote d’Azur, CNRS, Laboratoire Lagrange, Bd de l’Observatoire, F-06304 Nice, France\\
$^{2}$Universit\'e C\^ote d’Azur, Observatoire de la C\^ote d’Azur, CNRS, Artemis, Bd de l'Observatoire, F-06304 Nice, France\\
$^{3}$INFN, Sezione di Roma, I-00185 Roma, Italy}

\title[Progenitor formation galaxies of merging binary black holes]{Understanding the progenitor formation galaxies of merging binary black holes}
% \date{\today}

\maketitle
\begin{abstract}
With nearly a hundred gravitational wave detections, the origin of black hole mergers has become a key question. Here, we focus on understanding the typical galactic environment in which binary black hole mergers arise. To this end, we synthesize progenitors of binary black hole mergers as a function of the redshift of progenitor formation, present-day formation galaxy mass, and progenitor stellar metallicity for 240 star formation and binary evolution models. We provide guidelines to infer the formation galaxy properties and time of formation, highlighting the interplay between the star formation rate and the efficiency of forming merging binary black holes from binary stars, both of which strongly depend on metallicity. We find that across models, over 50\% of BBH mergers have a progenitor metallicity of a few tenths of Solar metallicity, however, inferring formation galaxy properties strongly depends on both the binary evolution model and global metallicity evolution. The numerous, low-mass black holes  ($\mathrm{\lesssim 15}$ \Msun) trace the bulk of the star formation in galaxies heavier than the Milky Way (\Mgal\ $\mathrm{\gtrsim 10^{10.5}}$ \Msun). In contrast, heavier BBH mergers typically stem from larger black holes forming in lower metallicity dwarf galaxies (\Mgal\ $\mathrm{\lesssim 10^{9}}$ \Msun). 
% In comparison to the astrophysical population, w
We find that the progenitors of detectable binary black holes tend to arise from dwarf galaxies at a lower formation redshift ($\lesssim$ 1). We also produce a posterior probability of the progenitor environment for any detected gravitational wave signal. For the massive GW150914 merger, we show that it likely came from a very low metallicity (\Z\ $\mathrm{\lesssim}$ 0.025 \Zsun) environment.
\end{abstract}

\section{Introduction}
\label{sec: Intro}
Gravitational wave (GW) astronomy has become a staple topic of discussion in the context of understanding the astrophysics and origins of compact objects like back holes (BHs) and neutron stars. The number of detections of binary black holes (BBHs) has exponentially risen from successive observing runs. In addition to the existing 90 \citep{GWTC-3} BBH detections declared by the LIGO-Virgo-KAGRA (LVK) collaboration, the advent of the fourth observing run (O4) is expected to bring the number of detections well into the hundreds \citep{ProspectsLVK}. With this wealth of GW data, we can statistically probe the origins of black holes by combining information from the detected GW parameters with progenitor star formation and evolution models.

Present-day GW detectors lack the sensitivity to identify the host galaxies of compact binaries due to their poor sky-localization \citep{ProspectsLVK}. The host galaxy of the first detected binary neutron star merger, GW170817 \citep{GW170817-Detection} was identified by observing its electromagnetic counterparts with telescopes with far greater angular resolution than GW detectors. Unfortunately, most BBH mergers are unlikely to have a detectable EM counterpart \citep{FermiEMFollowupmBBH}. Therefore, to understand their progenitor environment, we cannot rely on GW or electromagnetic observations alone and must include simulations to infer the formation galaxy parameters. Current literature has highlighted the challenge in predicting the progenitors of BBHs primarily due to uncertainties in modeling their formation mechanisms \citep[see][for recent reviews]{Mapelli2020_formchannel, MandelBroekgaardenReview2022}. Two widely accepted formation channels of merging BBHs are isolated binary evolution of black holes \citep[e.g.][]{Bethe+98_IsoEvol, Belczynski+2002, Kalogera+2007, Dominik+2012,Belczynski_16_GW150914,Eldridge_16_BBH,Stevenson_17_BBH,Kruckow_18_COMBINE,Giacobbo_18_BBH,Kruckow_18_COMBINE}, and dynamical interactions of black holes in dense environments such as young star clusters, globular clusters or nuclear star clusters which can lead to the formation of progressively heavier black holes through hierarchical mergers \citep[e.g.][]{Sigurdsson+93DynForm, Zwart+00_DynForm,Mapelli16_clusters,Rodriguez_15_BBH_clusters,Rodriguez_16_mergerrates,Askar_17_MOCCA,Samsing_18_clusters_ecc,McKernan_18_AGN_mergers,DiCarlo_19_YSC,Zevin_19_clusters_ecc, Mapelli2021_HierarchicalMergers}. In this paper, we focus on the former of the two mechanisms.

The astrophysical interpretation of GW events comes in different flavors. Some studies build detailed models for stellar and binary evolution which produce BBH mergers with global properties, such as masses and spins, that can be compared with observations \citep{Spera_19_SEVN,Marchant_21_MT,Bavera_21_MT,ZevinBavera_IsoEvolBHmassSpin,Bavera_30Mformation,Dorozsmai_23_Stable_MT}.  Other models rely on sets of simplified models, which can be easily used to build entire populations of compact objects, which can be directly compared to observations \citep{Mapelli2017mergerrate_Illustrius,GiacobboMapelli2018_efficiencyBBH,Mapelli_18_rates,Mapelli+2019_propmBBH,Baibhav+19_rates,Broekgaarden_21_BBH_MZR_vs_bin,Leike-Redshift-Evolution-mBBHR+2022}. These populations are based on a combination of models of binary evolution and metallicity-dependent global star formation in the Universe, which can be based on cosmological simulations, semi-analytic models and/or observations.  

Recent studies have highlighted the importance of the star formation model when it comes to population inference of merging black holes, as these often stem from low-metallicity progenitor stars. Modeling the whole range of progenitor metallicity across cosmic history is one key to understanding compact object mergers but it is far from straightforward \citep{Chruslinska+19}. One can assume a simple evolution with redshift \citep{GiacobboMapelli2018_efficiencyBBH}, including a scatter in the metallicity distribution at all redshifts \citep{Santoliquido_21_SFR_vs_binary}. Models based on cosmological simulations directly provide redshift-dependent distributions of the metallicity, across different  types of galaxies \citep{Mapelli2017mergerrate_Illustrius,Schneider+2017_hostgalaxy_firstGWs,Artale+2020_hostgalaxy}. Otherwise, one can use the combination of the mass-metallicity relation \citep[or the fundamental metallicity relation][]{Santoliquido_22_host_galaxies} across different galaxy masses with a galaxy stellar mass function to build a star formation model \citep{Boco2019_MergerRate,Neijssel+2020,Broekgaarden_21_BBH_MZR_vs_bin}. These studies find that the choice of star formation model, through the distribution of metallicity, is equally important as the choice of binary evolution model, when it comes to determining BBH merger rates.

Here, we study how the uncertainties on the metallicity distribution throughout cosmic history impact our inference on the typical progenitor environment of BBH mergers. To obtain a comprehensive understanding of the progenitor environment,  we investigate the progenitor properties along three dimensions: the present-day formation galaxy mass, the metallicity of the star-forming gas in the galaxy, and the progenitor formation time. At the heart of our model of the star formation history is the average star formation rate as a function of dark matter halo mass from \citet{behroozi2013}, calculated based on a self-consistent Markov Chain Monte Carlo algorithm that uses both star formation rate observations and dark-matter simulations. \citet{Neijssel+2020} analyzed the impact on merger rates from different galaxy stellar mass functions, metallicity models, and global star formation rates; demonstrating the choice of metallicity models has the strongest impact on the BBH merger rates. \citet{Chruslinska_19_influenceZ} have further highlighted the effect of observational uncertainties in the mass-metallicity relation, especially at high redshift ($>$ 3).  Building on this, we systematically consider multiple metallicity models of the star formation history and numerous binary evolution models in our analysis. 

Binary population synthesis codes are used to rapidly simulate the binary evolution of stars, allowing us to explore different models of binary evolution by varying flags within the code.
In this study, we perform the population synthesis using COSMIC \citet{Breivik:2019lmt} to study the formation conditions of progenitors of merging BBHs. Therefore, we take an agnostic approach where we explore popular models and focus on the presence of overlapping progenitor galaxy properties. 

In \S \ref{sec: Methods} we build a model of the progenitors of merging black holes along 3 dimensions of formation galaxy mass, metallicity, and time of progenitor formation by simulating the star formation history and binary evolution. Differentiating from other progenitor studies, we also simulate the effect of GW detection biases to contrast the progenitors of detected and astrophysical merging BBHs so as to better answer the question of where detected BBHs come from. Accounting for different models of the metallicity dependence of the star formation history and deviations from the default binary evolution model we explore the progenitor formation galaxy properties of merging BBHs in \S \ref{sec: Results}. Using our simulations, we describe a tool to produce progenitor posterior probabilities of the formation galaxy properties for individual GW detections (\S \ref{sec: Results-ProgenPosterior}). In \S\ref{sec: Discussion} we discuss the implications of our results and propose avenues for improving the interpretation of BBH mergers. 

\section{Methods}\label{sec: Methods}
\begin{figure}
    \centering
    \includegraphics[width=.45\textwidth]{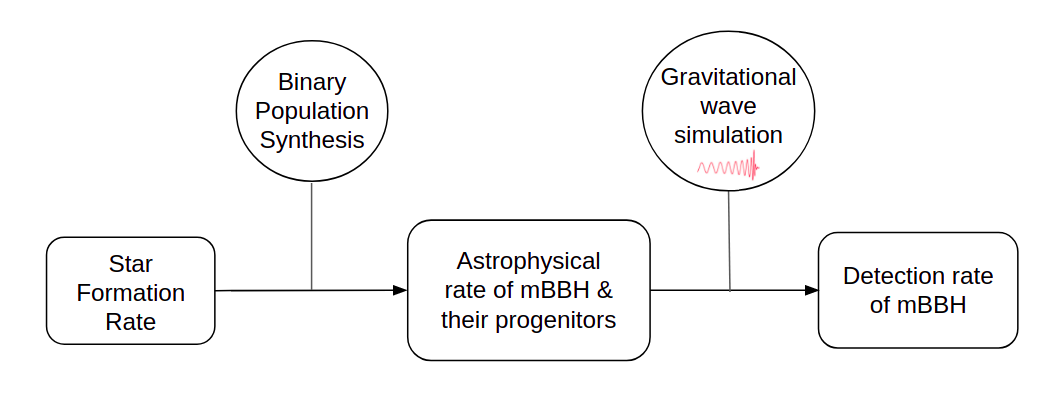}
    \caption{Workflow of our 
    pipeline to simulate the merging BBH population. First, we model the star formation rate as a function of metallicity, galaxy mass, and redshift of formation. Then, the astrophysical formation rate is obtained by integrating the star formation rates with the binary population synthesis of binary stars that form merging binary black holes. The detection rate of binary black holes is calculated by simulating the detection process of gravitational waves emitted by the astrophysical population.}
    \label{fig: flowchart}
\end{figure}

To understand the progenitor environment of BBHs that merge by the present-day, hereafter referred to as mBBHs, we create a set of models that combine star formation models and binary stellar evolution models. We parametrize the star formation environment by the present-day formation galaxy mass \Mgal, the metallicity of star-forming gas in the galaxy \Z, and the redshift of star formation \zf\ (\S\ref{sec: SFR}). Using a binary population synthesis code,  we produce a representative population of mBBHs from binary systems with different metallicities (\S\ref{sec: BPS}). We then convolve the star formation rate with the binary models to get the astrophysical merger rate of mBBHs and the formation rate of their progenitors (\S\ref{sec: astro progen rate} and \S\ref{sec: astro merger rate}). Finally, we generate the GW signals emitted by these systems and simulate their detection by the LIGO-Virgo three-detector network considering their sensitivity during the third observing run (\S\ref{sec: Methods- detectable merger rate}). We describe the construction of our simulation in the flowchart shown in Fig. \ref{fig: flowchart}.

\subsection{Star Formation Rate}
\label{sec: SFR}

\begin{figure*}
    \centering
    \includegraphics[width=\linewidth]{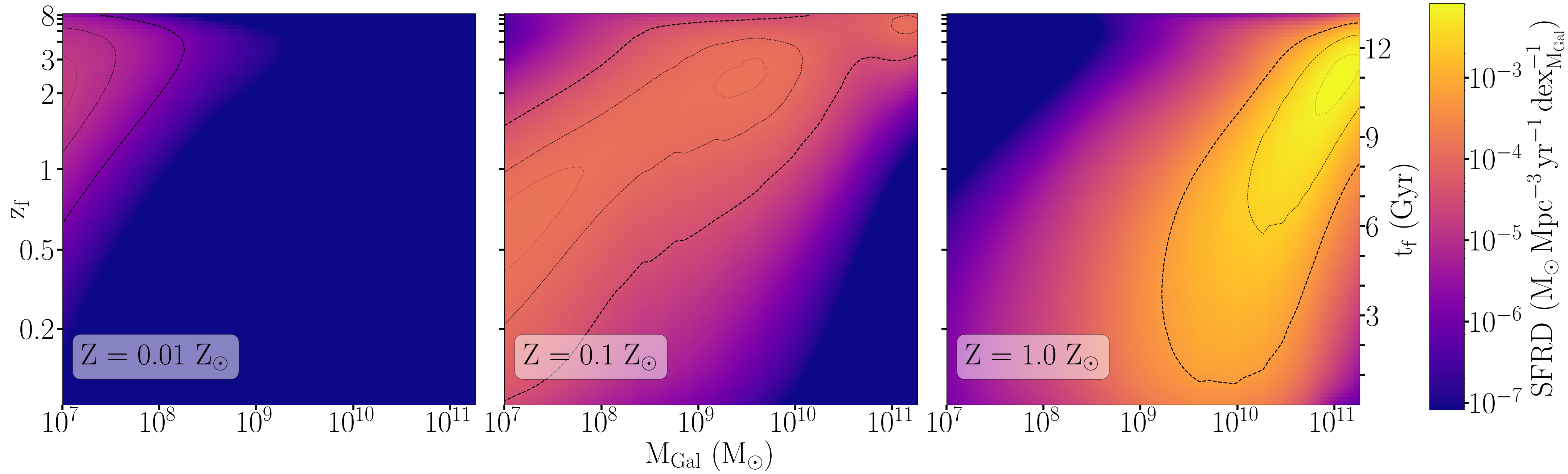}
    \caption{Star formation history for different metallicities in our default model. We show the star formation rate density (SFRD), depicted by the logarithmic color map, as a function of present-day galaxy mass \Mgal\ and formation redshift \zf\ or look-back time \tf\ for metallicities \unit[0.01]{\Zsun} (left), \unit[0.1]{\Zsun} (center) and \unit[1.0]{\Zsun} (right). The contour lines, in decreasing order of line width, contain the $\mathrm{90^{th}}$, $\mathrm{50^{th}}$ and $\mathrm{10^{th}}$ percentile of the SFRD. This particular SFRD is generated using the mass-metallicity relation from the \citet{Ma+2016} fit for the \citet{KK04} calibration (KK04\_Ma16).}
    \label{fig: SFR3D_KK04_Ma16}
\end{figure*}

\begin{figure*}%[H]
    \centering
    \includegraphics[width=\linewidth]{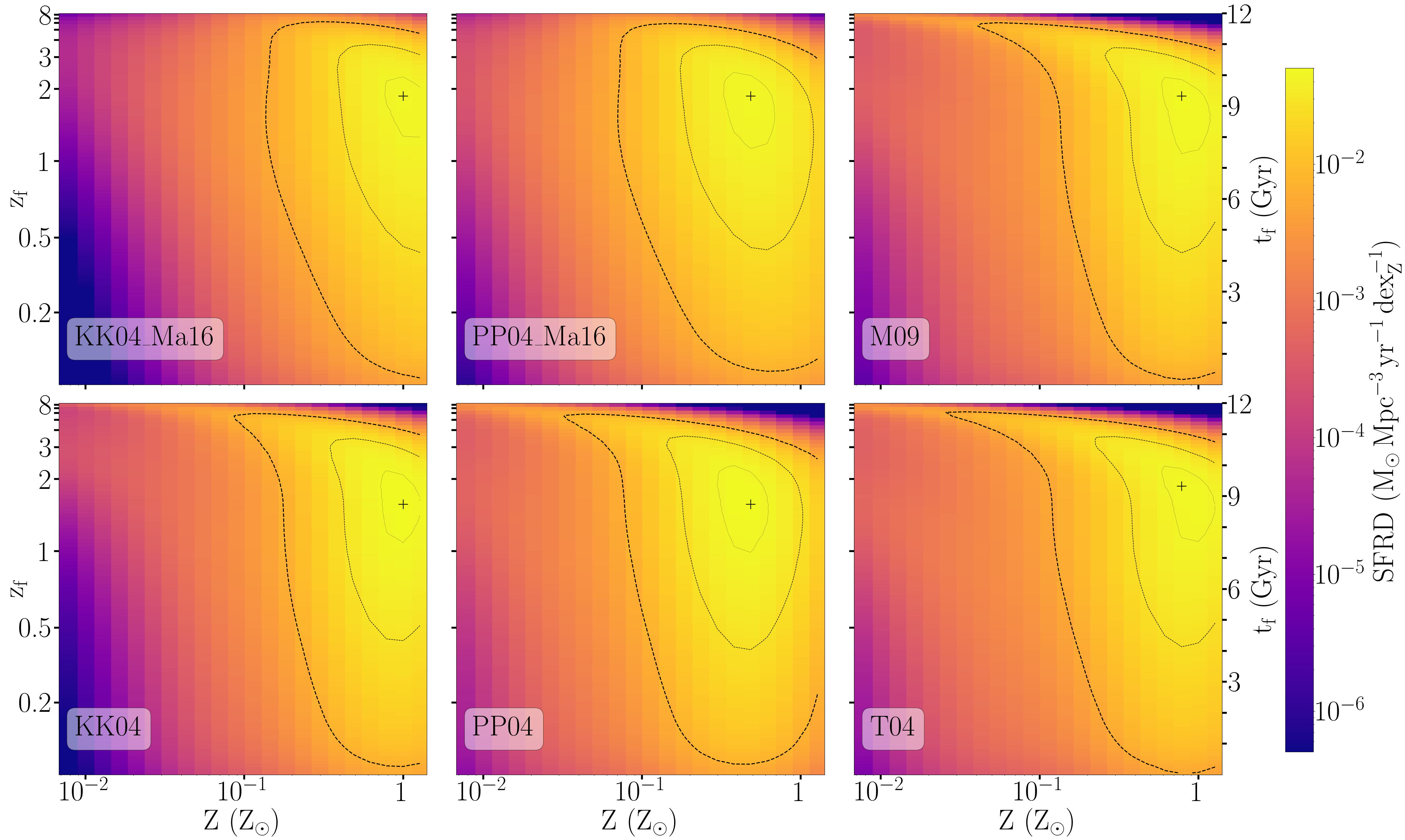}
    \caption{Star formation history as a function of metallicity for the different mass-metallicity models that we study. Clockwise from the top-left, we show the star formation rate density, depicted by the logarithmic color map, based on different mass-metallicity relations: \citet{Ma+2016} fit for \citet{KK04} (KK04\_Ma16),  \citet{Ma+2016} fit for \citet{PP04} (PP04\_Ma16), \citet{Chruslinska+19} fit based on \citet{M09} refinement of \citet{Maiolino+08} (M09), \citet{Chruslinska+19} fit for \citet{T04} (T04), \citet{Chruslinska+19} fit for \citet{PP04} (PP04) and \citet{Chruslinska+19} fit for \citet{KK04} (KK04). The contour lines in decreasing order of line width, contain the $\mathrm{90^{th}}$, $\mathrm{50^{th}}$ and $\mathrm{10^{th}}$ percentile of the SFRD, with the peak shown by '+'.}
    \label{fig: SFR_Z_z}
\end{figure*}

The progenitor formation rates of mBBH depend on the star formation in the Universe and the efficiency of stars that evolve in binary systems ultimately forming  mBBHs. The star formation rate density (SFRD) describes the mass time-volume rate of star formation in the Universe.
Because the mass of the formation galaxy influences the amount of star-forming gas available to form binary systems, we need to consider the SFRD as a function of \Mgal\ and \zf. Moreover, the metallicity of the star-forming gas is correlated with \Mgal\ and also strongly influences the evolution of the binary star systems. Thus, for the purpose of producing astrophysical merger rates, we are interested in building a 3-dimensional model of the SFRD as a function of \Z, \Mgal, and \zf. To this end, we follow a modified version of \citet{Lamberts2016}.

To develop the 3-dimensional SFRD, we first use the star formation rate (SFR) from \citet{behroozi2013} which provides the average observed SFR (in units \Msun\ $\mathrm{yr^{-1}}$) as a function of present-day dark matter halo mass and redshift \zf.
The mapping between the present-day halo mass and its galaxy mass is provided by the abundance-matching technique described in \citet{abundance_matching:Behroozi_2010}. It maps the rank order galaxy mass with the rank order halo mass such that the heaviest halos contain the heaviest galaxies and vice-versa. To account for the occurrence of different galaxy masses, we use the galaxy stellar-mass function $\frac{\mathrm{d}^2N_\mathrm{Gal}(M_\mathrm{Gal})}{\mathrm{dlog}\,M_\mathrm{Gal}\, \mathrm{d}V_\mathrm{c}}$ derived from observations that describe the number density of galaxies as function of \Mgal. For our analysis, we use the galaxy stellar-mass function fitted by \citet{Tomczak+2014}.
To model the metallicity of star-forming gas in galaxies, we consider the mass-metallicity relation (MZR) that describes the mean metallicity of the star-forming gas as a function of \Mgal\ and \zf. The metallicity \Z\ is defined as the mass fraction of elements heavier than helium. We use the value of the metallicity relative to that of the Sun,  $Z \mathrm{\coloneqq [Z/H]\,=\,log_{10}[(Z/H)/(Z/H)_{\odot}]}$ and we assume that the value of Solar metallicity \Zsun\ = 0.014 \citep{Asplund+2009}.
Observationally, it is often easier to measure the abundance of certain elements such as oxygen.
The abundance ratio of other elements are usually assumed to scale linearly with that of oxygen, maintaining the solar abundance ratios.
Several  values are available across the  literature \citep[see e.g.][]{Chruslinska+19}, and we assume the solar oxygen abundance follows 12 + $\mathrm{log_{10}(O/H)_{\odot}}$ = 9. 

Modeling the MZR strongly depends on the method used to approximate (O/H). Accounting for variations in MZR modeling, we explore different calibrations used to estimate (O/H) and different fits used to extrapolate the MZR to high redshifts. The direct method of estimating (O/H) involves measuring auroral lines from the H{\small II}-regions. However, these emission lines are typically weak. Instead, several calibrations have been used to estimate the oxygen abundance from stronger emission lines. Popular calibrations include \citet{PP04}, which use measurements of the gas electron temperature, \citet{KK04} and \citet{T04}, which are based on the photo-ionization mechanism, and large redshift measurements from deep near-IR spectroscopy as done in \citet{Maiolino+08}, later refined by \citet{M09}. 

Moreover, at high redshifts (\zf $\mathrm{\gtrsim}$ 3), measurements can be sparse and different observational techniques must be used, such as Damped Lyman $\alpha$ (DLA) measurements \citep{DLAs-Rafelski}. Therefore, connecting low and high redshift measurements of the MZRs can be challenging. We use two fits of the MZR with significantly different models at high redshifts. The first fit is prescribed in Eq. 3, 4 and Table 2 of \citet{Chruslinska+19} for the aforementioned metallicity calibrations (whose MZRs will be henceforth denoted as PP04, KK04, T04, and M09 respectively). Our second fit is provided by \citet{Ma+2016} for the \citet{PP04} and \citet{KK04} calibrations (MZRs denoted as PP04\_Ma16 and KK04\_Ma16 respectively) which are based on high-resolution cosmological zoom-in simulations of galaxy formation using the FIRE suite \citep{Hopkins+2014} and consistent with observations at low redshift. Due to the large uncertainties in the calibrations and extrapolation to higher redshift, we explore all six MZRs in our analysis. By default, we use the MZR of KK04\_Ma16 as it shows reasonable consistency with observations at low redshift.

Accounting for variance around the mean metallicity, we approximate the metallicity distribution in a galaxy as a Normal distribution $\mathcal{N}$ with a mean metallicity from the MZR and a scatter \scatter\ of three origins: $\simeq$ \unit[0.1]{dex} due to differences between galaxies \citep{T04}, $\simeq$ \unit[0.2]{dex} due to radial variations within a galaxy \citep{Henry+2010} and $\simeq$ \unit[0.2]{dex} for differences at a given galactocentric radius \citep{Berg+2013}. Adding the contributions in quadrature, we get \scatter\ = \unit[0.3]{dex}. This scatter accounts for the large variance in metallicity seen within galaxies, including the lower metallicity seen in the outskirts of galaxies \citep{Chakrabarti_2017_alternateChannel} which provides an ideal environment for BBH formation. Thus, the fraction $\Psi$(\Z,\Mgal,\zf) of stars forming within a metallicity bin
is the integral of $\mathcal{N}(\mu\mathrm{=12+log_{10}}(O/H),\,\sigma\mathrm{=0.3})$ over the metallicity bin \Z\ which corresponds to the metallicity bins used to generate the binary population in \S \ref{sec: BPS}.

We derive the SFRD, $\frac{\mathrm{d}^4M_*}{\mathrm{d}V_{\mathrm{c}}\, \mathrm{d}t_{\mathrm{f}}\, \mathrm{dlog}M_{\mathrm{Gal}} \,\mathrm{dlog}Z}$, by incorporating the star formation history, the galaxy-mass to dark matter halo-mass abundance matching, the stellar mass function, and the metallicity dependence of star formation 

\begin{eqnarray}
\frac{\mathrm{d}^4M_*(Z, M_{\mathrm{Gal}}, z_{\mathrm{f}})}{\mathrm{d}V_{\mathrm{c}}\, \mathrm{d}t_{\mathrm{f}}\, \mathrm{dlog}M_{\mathrm{Gal}} \,\mathrm{dlog}Z}&=&\frac{ \mathrm{d}M_*(M_{\mathrm{Gal}},z_{\mathrm{f}})}{\mathrm{d}t_{\mathrm{f}}}\\
&\times& \frac{\mathrm{d}^2 N_\mathrm{Gal}(M_{\mathrm{Gal}})}{\mathrm{dlog}M_{\mathrm{Gal}}\,\mathrm{d}V_{\mathrm{c}}} \times \frac{\mathrm{d}\Psi(Z,M_{\mathrm{Gal}},z_{\mathrm{f}})}{\mathrm{dlog}Z} \nonumber
\label{eqn: SFRD}
\end{eqnarray}
where $\frac{\mathrm{d}M_*(M_\mathrm{Gal},z_\mathrm{f})}{\mathrm{d}t_\mathrm{f}}$ is the star formation history of a galaxy with a present-day mass \Mgal. Following the limits on \Mgal\ and \zf\ from \citet{behroozi2013}, we span the ranges \unit[$10^{7} - 10^{11.25}$]{\Msun} and $0-8$ respectively. The range of metallicity covered by the binary population synthesis code is from $\mathrm{10^{-4}}$ to $\mathrm{1.78\times10^{-2}}$. We use these ranges for our global model of merging black holes. Upon checking for robustness to different bin-sizes, the galaxy-mass is logarithmically spaced with 200 samples, the redshift of formation is distributed uniformly in look-back time in intervals of \unit[100]{Myr}, and the metallicity bins are logarithmically spaced with 22 bins.

We show the SFRD as a function of \Mgal, and \zf\ in Fig. \ref{fig: SFR3D_KK04_Ma16} for the KK04\_Ma16 model. Globally, the peak of star formation occurs at \zf\ $\simeq$ 2.5, in galaxies larger than the Milky-Way (\Mgal\ $\simeq$ \unit[$10^{11}$]{\Msun}) at nearly solar metallicity. Looking at star formation at different metallicity ranges, very low metallicity stars (\Z\ $\simeq$ \unit[0.01]{\Zsun}) are few in number and tend to form in dwarf galaxies (\Mgal\ $\lesssim$ \unit[$10^9$]{\Msun}) at high redshifts (2 < \zf\ < 4). In contrast, high metallicity (\Z\ $\simeq$ \unit[]{\Zsun}) stars are more numerous and are predominantly formed in large galaxies (\Mgal\ $\gtrsim$ \unit[$10^{10}$]{\Msun}) over a wide range of redshifts. Stars with metallicity \Z\ $\simeq$ \unit[0.1]{\Zsun} can form in all types of galaxies. They are formed in dwarf galaxies at low redshift (\zf\ < 1), average sized galaxies ($\mathrm{10^{9}\,to\,10^{10}}$ \Msun) at high redshifts (2 < \zf\ <4), and large galaxies at very high redshifts (\zf\ > 4).

Fig. \ref{fig: SFR_Z_z} shows the SFRD as a function of \Z, and \zf\ for the six different SFR/MZR models. The distributions show different peak values in terms of metallicity. The KK04 calibration has the highest mode of metallicity (\Z\ $\mathrm{\simeq}$ \Zsun), followed by mode of the M09 and T04 calibrations. The PP04 calibration shows the lowest mode of metallicity (\Z\ $\mathrm{\simeq}$ 0.5).
Comparing the 90\% contours of the SFRD in different extrapolated fits of the MZR at high redshift (\zf\ $\mathrm{\gtrsim 3}$), we see that the Ma\_16 fits (corresponding to SFR/MZR of KK04\_Ma16 and PP04\_Ma16) show most stars are formed with \Z\ $\mathrm{\gtrsim}$ \unit[0.1]{\Zsun}. In contrast, fits which consider a smooth drop in the MZR with redshift (corresponding to SFR/MZR of KK04, PP04, T04 and M09) predict that, for $z_f>3$, most stars are formed with \Z\ $\mathrm{\lesssim}$ 0.1 \Zsun.

As we are interested in binary systems, we include the binary fraction, the fraction of binaries among stellar systems, into our SFRD. It is set to a constant value of 0.7 based on observations of massive stars \citep{sana12}.

\subsection{Binary Population Synthesis}
\label{sec: BPS}

\begin{figure}
    \centering
    \includegraphics[width=\columnwidth]{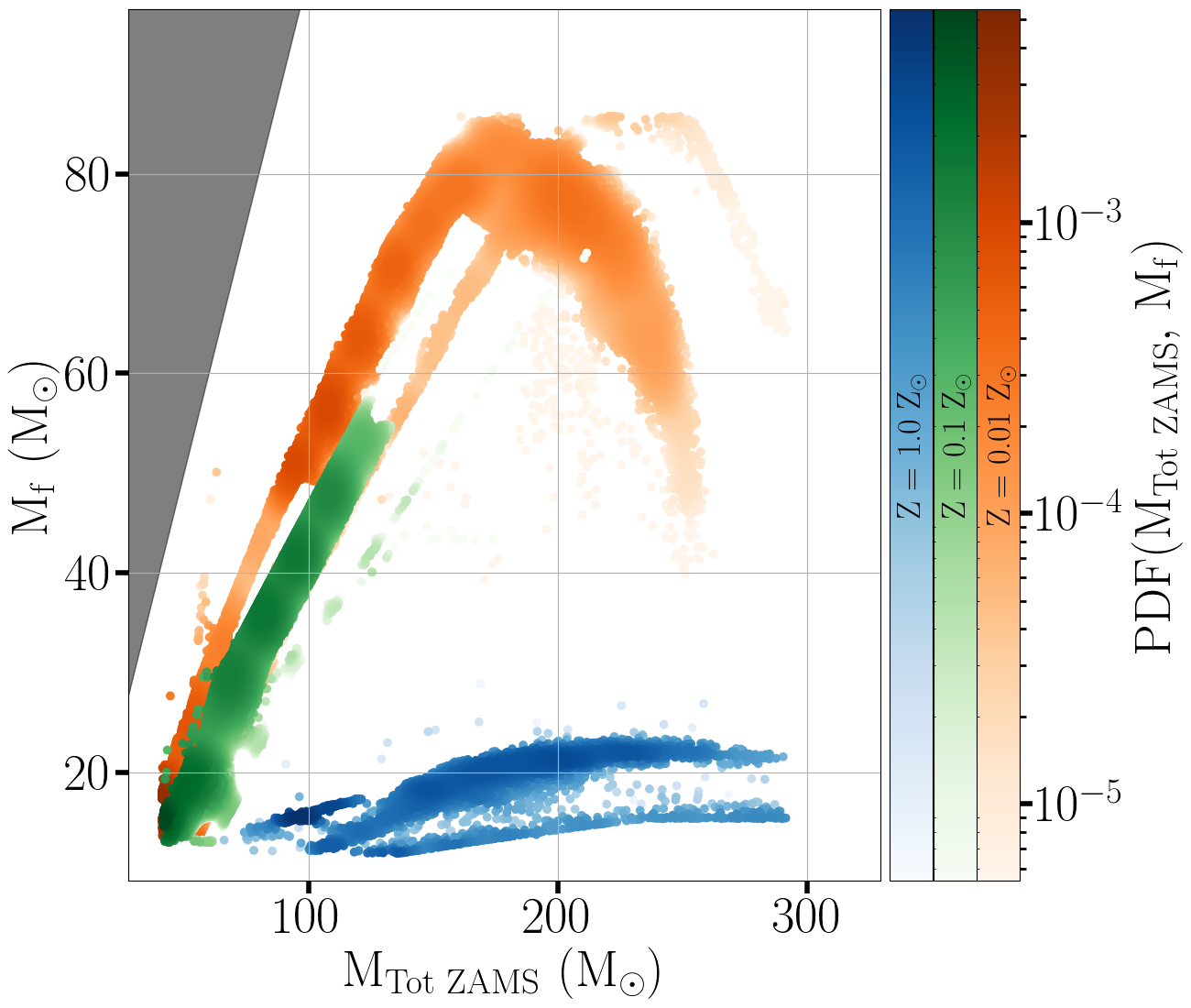}
    \caption{Remnant black hole mass $M_\mathrm{f}$ vs total initial stellar mass $M_\mathrm{Tot\,\,ZAMS}$ from our default binary population synthesis simulation for metallicities 0.01 \Zsun\ (red), 0.1 \Zsun\ (green) and 1.0 \Zsun\ (blue). The color map represents the probability density function for each metallicity.}
    \label{fig: BPS}
\end{figure}

To generate mBBHs, we use COSMIC \citep{Breivik:2019lmt} which is a rapid binary population synthesis code that builds on the population synthesis code from \citet{BSEHurley2002}. From this catalog, we identify a representative sub-population of binary star systems with varying metallicities \Z\ and binary parameters $\theta_\mathrm{BPS}$ that ultimately form mBBHs, denoted by \PopBPS(\Z, $\theta_\mathrm{BPS}$).

For every metallicity, \Z, we use COSMIC to iteratively sample and evolve binary zero-age main sequence (ZAMS) stars, selecting those that form mBBHs until recovering a converging distribution of BBH stellar masses and orbital parameters at the time of formation with a tolerance mismatch of at most $\mathrm{10^{-4}}$. The primary star mass is sampled from the initial mass function in \citet{kroupa01} between 15 and 100 \Msun and the secondary mass is sampled from a uniform mass distribution from 15 \Msun\ to primary mass. The orbital parameters are sampled from \citet{sana12}. For each BPS model, across all metallicities, we evolve $1.2\times 10^{11}$ \Msun\ binary ZAMS star masses, forming a converging population of $2.2 \times 10^6$ progenitors of mBBHs.

The evolution of massive binary stars depends on many parameters describing both the evolution of singles stars and their binary interactions. The formation of mBBH systems is mostly affected by the choice of model parameters for stellar winds, mass transfers, supernova mechanisms, and kicks. The binary evolution models for these parameters are categorized as flags in COSMIC and, in this study, our default values have been set to values prescribed by the COSMIC documentation, v3.4.0
\footnote{https://github.com/COSMIC-PopSynth/COSMIC/tree/v3.4.0}.
In the following paragraphs, we describe relevant flags and the values we explore. When exploring different values of a given parameter, all other parameters are set to their default values, thus varying only one parameter at a time. Note that we are not interested in commenting on the validity of the parameter values. Instead, we include variations in the BPS modeling to infer and compare the progenitor environments they predict. Table \ref{tab:BPS flag} of the Appendix \ref{sec: Appendix} summarizes the values and corresponding BPS model of the parameters we explore.

Winds are expulsions of stellar material, primarily due to radiation pressure. Winds increase with increasing metallicity, due to enhanced opacity in the stellar atmosphere  \citep{Vink+2001, Vink+2005}. The wind parameters we explore include the wind velocity factor ($\beta_\mathrm{w}$) that relates the wind velocity to the surface escape velocity of the star. $\beta_\mathrm{w}$ is set to the \citet{Belczynski+2008} prescription by default and we explore high (7) and low (0.125) values for Eq. 9 of \cite{BSEHurley2002}. The $\dot{M}L_\mathrm{edd}$ flag, which describes the mass loss near the Eddington limit, is set to be independent of the metallicity by default and we explore the metallicity dependence from \citet{GiacobboMapelli2018_efficiencyBBH}.

Mass-transfers can greatly influence the orbital separation of the binary \citep{alphalambdaFormalism, Mapelli2018review}. Typically, a mass transfer is initiated when the stellar radius of one star (donor) exceeds its Roche lobe which results in a mass transfer from the donor to the accretor that can modify the orbital separation. It starts off as a stable mass-transfers and if the donor is unable to adjust to the mass transfer, a runaway process begins that occurs on a dynamical timescale. This results in a common envelope (CE) surrounding the binary \citep{ivanova2013CE}. This is a non-conservative process leading to a significant loss in mass and orbital momentum of the system, thereby shortening the orbital separation until the envelope is ultimately ejected or the binary prematurely merges prior to BBH formation. 

We explore different values of  the CE efficiency of transferring orbital energy to the envelope ($\alpha$). We set the default, to 1 and we explore the values 0.1, 0.2, 0.3, 0.5, 2, 3, 5, 10. Our default value for the binding energy factor of the envelope $\lambda$ is set to the prescription from the appendix of \citet{Claeys+2014}, without the extra ionization energy. We also explore a case with the extra ionization energy as well as a constant $\lambda$ set to $1/\alpha$ (see \S 2.7.1 of \citet{BSEHurley2002}). We also check a CE prescription that prevents the direct stellar merger of systems lacking a core-envelope boundary of the donor ($CE_\mathrm{Merger}$(0)). Regarding the initial orbital energy calculation ($E_\mathrm{orb, i}$), we use, by default, the core masses as per Eq. 70 of \citet{BSEHurley2002} and we explore the calculation with core + envelope mass from the \citet{deKool+1990} prescription. The critical mass ratio model ($q_\mathrm{crit}$) determining the onset of CE is given by Table 2 of \citet{Claeys+2014} and we explore alternate models from \citet{BSEHurley2002}, \citet{Belczynski+2008}, and \citet{Hjellming_Webbink_qcrit_1and3} for GB/AGB stars.

Supernova kicks can also impact the final fate of the BBH by modifying its orbit and potentially unbinding the binary. These natal kicks are imparted due to the conservation of momentum of the binary due to the mass lost as ejecta \citep{Blaauw} and possibly because of asymmetric stellar collapse. In our default model, we use the natal kick prescription $v_\mathrm{Kick}$ of \citet{BSEHurley2002} which is based on pulsar observations, where kicks are drawn from a Maxwellian distribution with dispersion parameter, $\sigma$, set to 265 km/s. We also simulate large (530 km/s) and small (90 km/s) values of $\sigma$. We explore other kick models described by two equations from \citet{GiacobboMapelli2020}, and a relation for neutron star kick velocity from \citet{BrayEldridge2016}. Eq 1 of \citet{GiacobboMapelli2020} scales the sampled kick velocity by the relative ratio of the ejecta mass by the remnant mass with respect to that of the typical neutron star and ejecta masses. Eq 2 scales the velocity by the ratio of the ejected mass with the typical ejecta mass for neutron star formation.
\citet{BrayEldridge2016} describes the kick as a linear function of the ratio of the ejected mass to remnant mass. Upon sampling the natal kick using $v_\mathrm{Kick}$ and $\sigma$, it is modulated for BHs by the $v_\mathrm{BH}$ flag. By default, the \citet{Fryer+2012} prescription accounts for ejecta fallback. We also explore a model that re-weights the natal kick by the ratio of the remnant black hole mass to that of- a typical neutron star (1.44 \Msun), and a model that does not alter the sampled natal kick. 

The prescription of the supernova can influence the mass of the remnant drastically, denoted by the $M_\mathrm{NS-BH}$ flag. We set our default supernova mechanism to the rapid prescription of \citet{Fryer+2012} with the proto-core mass from \citet{GiacobboMapelli2020} which results in a mass-gap between neutron stars and black holes. We also explore \citet{Belczynski+2008}, and the delayed prescription from \citet{Fryer+2012}, both of which fill the mass-gap. The pair-instability supernova (PISN) and the pulsational pair-instability supernova (PPISN) mechanisms are modeled by the $PISN$ flags. The default flag uses the \citet{Marchant+2018} prescription and we explore the models of \citet{SperaMapelli2017} and \citet{Woosley2019}. In addition, we also switch off the PISN and PPISN mechanisms. Finally, we also explore the effect of different tide modeling, $Tide_\mathrm{ST}$, using \citet{Belczynski+2008} for the default and exploring \citet{BSEHurley2002}.

Fig. \ref{fig: BPS} shows the relation between the total ZAMS mass and the mass of the post-merger remnant black hole for different metallicities using our default BPS flags. High metallicity binaries tend to have light pre-supernova stars which can hamper BH formation or result in small BHs. This is due to high winds causing large mass loss during the evolution of the binary. The low-mass BHs are subject to large BH kicks, potentially disrupting the binary. Irrespective of metallicity, in this model, there appear to be two channels of binary evolution based on whether the binary was formed with equal mass or not. Systems with nearly equal masses have similar stellar evolution timescales. During their evolution, the binaries enter giant phases together, where both stars undergo a common envelope phase where their respective stellar envelopes are stripped, leaving behind a naked Helium core. The double CE event results in significant mass loss. Hence, equal mass ZAMS tend to form smaller remnants. In contrast, unequal mass ZAMS do not necessarily both undergo CE, thus resulting in lower mass loss and larger remnants. 

The drop in final remnant mass for total ZAMS mass $\mathrm{\gtrsim}$ 170 \Msun\ is due to our default PISN prescription. As shown in Table 1 of \citet{Marchant+2018}, as the pre-supernova mass of the progenitor stars increases beyond $\mathrm{\gtrsim}$ 50 \Msun, there is significant mass loss due to the PPISN mechanism resulting in a black hole with progressively smaller mass.

The synthetic catalogs of binary stars allow us to compute the efficiency of forming mBBHs from binary stars, defined as the number of mBBH progenitors per unit mass of initial binary stars.
\begin{equation}
\begin{aligned}
% \centering
    \eta_\mathrm{mBBH}(Z)=\frac{n_\mathrm{BPS}(Z)}{M_\mathrm{BPS}(Z)},
\end{aligned}
\end{equation}
where \MassPopBPS\ is the total initial stellar mass sampled to produce \PopBPS\ and \lenPopBPS\ is the number of mergers.

We split the range of metallicities given in \S \ref{sec: SFR} in 22 log-spaced bins, optimized for computational efficiency and checking for convergence in the \PopBPS\ population to different binning.

\subsection{Merging binary black hole population}

\subsubsection{Astrophysical mBBH progenitor population and formation rates}
\label{sec: astro progen rate}

We produce an astrophysical population of mBBH progenitors, \PopAstro, by randomly sampling a fixed number of systems, $N_\mathrm{s}$, from \PopBPS\~for every element in the 3D progenitor parameter space (\Z, \Mgal, \zf). We choose $N_\mathrm{s}=100$ as it is a good compromise between population robustness, computational efficiency, and memory management. We discard BBHs that merge after the present day. Every sampled system is assigned an astrophysical progenitor formation rate that is proportional to the SFRD and the efficiency of mBBH formation $\eta_\mathrm{mBBH}$.

\begin{eqnarray}
% \begin{aligned}
    R_\mathrm{i}(Z,M_\mathrm{Gal},z_\mathrm{f})&=&\frac{\mathrm{d}^2 N_\mathrm{i}(Z,M_\mathrm{Gal},z_\mathrm{f})}{\mathrm{d}V_\mathrm{c} \,\mathrm{d}t} \\
    &=& \frac{\mathrm{d}^2 M_\mathrm{*\,i}(Z,M_\mathrm{Gal},z_\mathrm{f})}{\mathrm{d}V_c\, \mathrm{d}t} \times \eta_\mathrm{mBBH}(Z) \times \frac{1}{N_\mathrm{s}}, \nonumber
% \end{aligned}
\label{eqn: PopAstro rate}
\end{eqnarray}

where $i$ denotes the $i^\mathrm{th}$ mBBH progenitor system in \PopAstro\ and $\mathrm{d^2}N_\mathrm{i}$ is the number of such systems within the differential co-moving time-volume element d$V_\mathrm{c}\,\mathrm{d}t$. $\frac{\mathrm{d^2}M_\mathrm{*\,i}}{\mathrm{d}V_\mathrm{c}\mathrm{d}t}$ is the SFRD from Eq \ref{eqn: SFRD} integrated over the width of the galaxy mass and metallicity bin. $R_\mathrm{i}$ is expressed in units of $\mathrm{Mpc^{-3} yr^{-1}}$. We compute the 3D astrophysical rate of mBBH progenitor formation, $R$, by summing over all mBBH progenitor systems.

\subsubsection{Astrophysical merger rate}
\label{sec: astro merger rate}
The merger rate of systems that coalesce within a look-back time bin centered at \tmbbh\ can be directly mapped back to the corresponding progenitor formation rate within an identically sized time bin centered at \tf~ \citep{Dominik2013}, with \Deltat\ = \unit[100]{Myr}. We obtain the merger rate evolution, $R_\mathrm{mBBH}$(\zmbbh), by summing over the progenitor parameter space the product of the progenitor formation rate and the corresponding fraction of progenitors that merge within the time bin \tmbbh.

\subsubsection{Progenitors of detectable mBBHs}
\label{sec: Methods- detectable merger rate}
To understand the detector selection effect on the progenitor population of mBBHs, we simulate GWs from the merger of every mBBH system in \PopAstro. We use the phenomenological inspiral-merger-ringdown waveform approximant IMRPhenomD \citep{Khan:2015jqa}, implemented using the python library PyCBC \citep{Usman:2015kfa}. The masses and \zmbbh\ of the merging black holes are extracted from our astrophysical population \PopAstro. The luminosity distance is calculated from \zmbbh\ using a flat $\Lambda$CDM cosmology model with the Plank 2016 cosmological parameters $\Omega_\mathrm{M}$ = 0.308, $\Omega_\mathrm{\Lambda}$ = 0.691, and Hubble constant $H_\mathrm{0}$ = 67.7 km/s/Mpc \citep{Plank+2016}. We assume the black holes are non-spinning. We distribute the sky positions isotropically and we sample the cosine of the inclination angle from a uniform distribution.

We consider a GW detector network of LIGO Hanford (H), LIGO Livingston (L) and Virgo (V) with their sensitivity during the second half of the O3 observation run \citep{GWTC-3}. We designate a binary BH coalescence as detectable if the signal-to-noise ratio in each detector is larger than 6 and the network signal-to-noise ratio is larger than 12. 

\section{Results}

\begin{figure*}%[H]
    \centering
    \includegraphics[width=\textwidth]{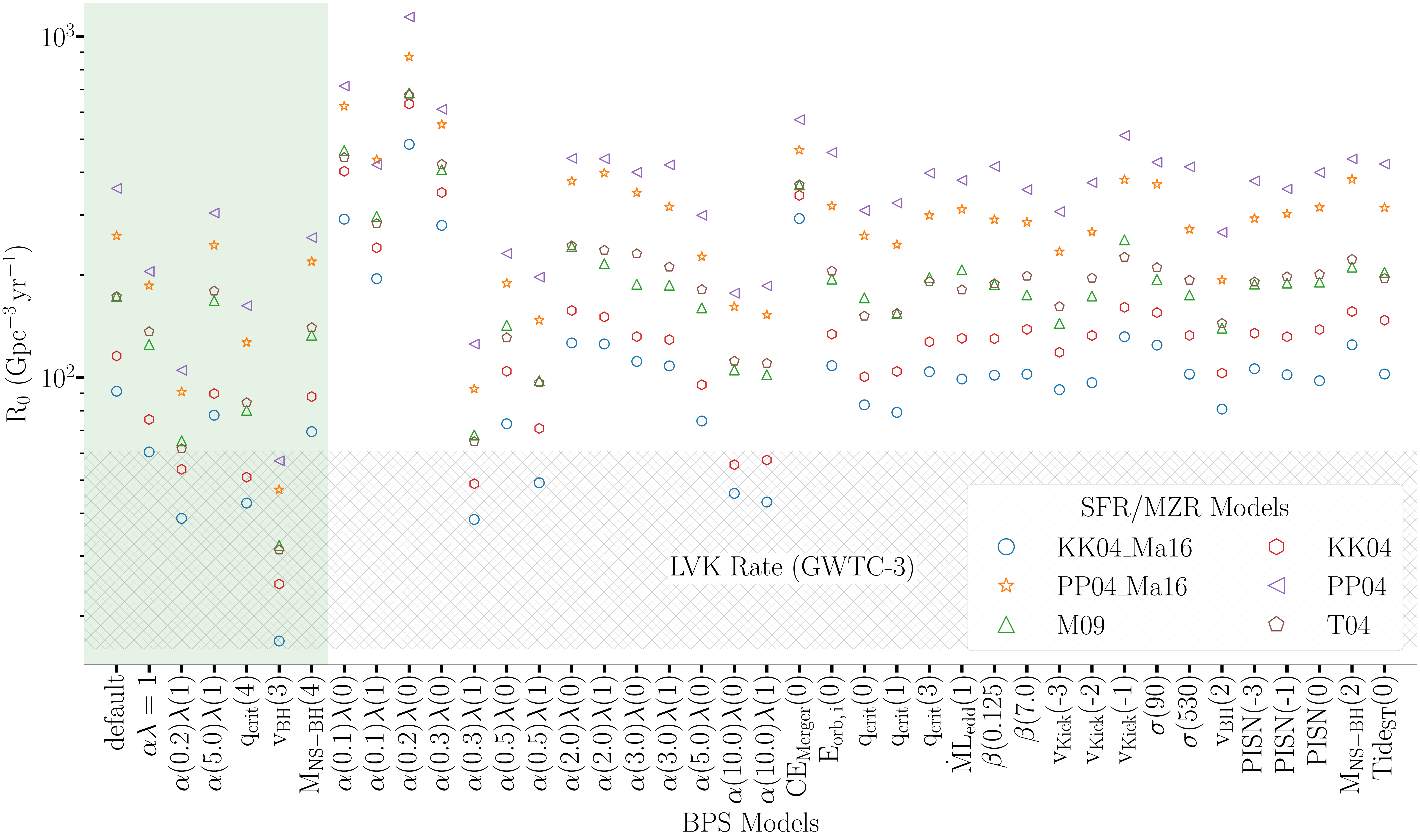}
    \caption{Present-day astrophysical merger rate density for our different models. We show the merger rate in our binary population models along the x-axis and the different mass-metallicity models of the star formation rate are shown with different colors and symbols (legend). The hatched region designates LVK's 90\% confidence interval of the inferred merger rate \citep{GWTC-3} after O3. The BPS models shaded in green will be further explored in later sections.}
    \label{fig: R0_BPSxCalib}
\end{figure*}

\label{sec: Results}
To understand the effect of different models in predicting the progenitors of mBBHs, we generate 240 models from a combination of the 40 BPS models and the SFRs based on the 6 mass-metallicity relations. First, we discuss the present-day astrophysical merger rate of the different models and how it connects with the efficiency of forming mBBHs as a function of metallicity and the global metallicity distribution in the Universe (\S\ref{sec: Results-AstroMergerRates}). Then, in \S\ref{sec: Results-AstroProgen}, we show the progenitor environment predicted by the different combinations of models and highlight their main differences and common points. In \S\ref{sec: Results-DetProgen}, we look at the selection bias introduced by GW detectors on the detected mBBH population and their inferred progenitor environment. Finally, in \S\ref{sec: Results-ProgenPosterior}, we describe a novel tool to produce posteriors of the formation galaxy properties for a given GW detection, taking GW150914 as an example.

\subsection{Astrophysical merger rates}
\label{sec: Results-AstroMergerRates}

\begin{figure*}%[H]
    \centering
    \includegraphics[width=\textwidth]{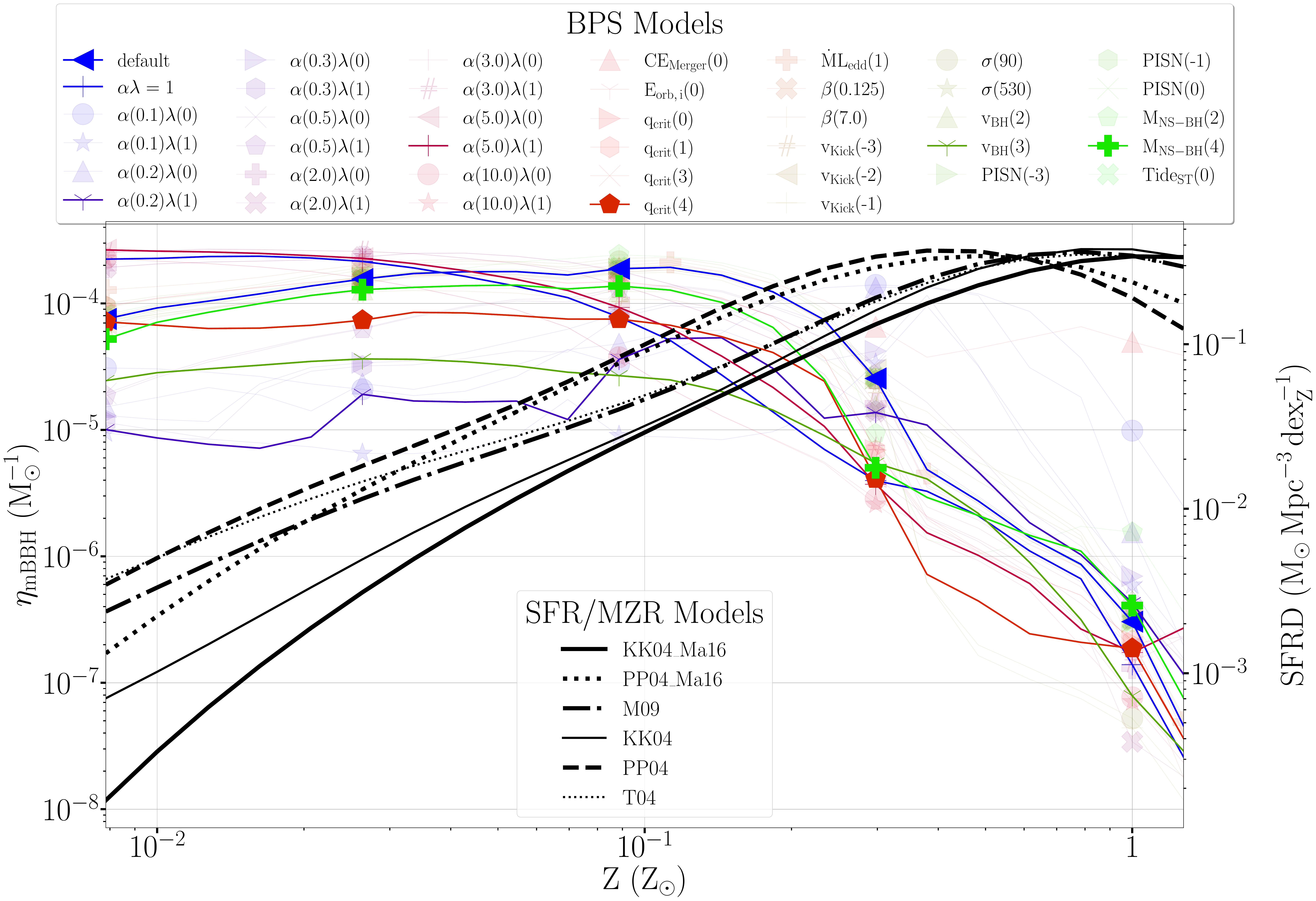}
    \caption{Efficiency of mBBH formation, $\eta_\mathrm{mBBH}$, as a function of metallicity \Z. We show all the binary population models (upper legend) and highlight the models which are chosen for further analysis with brighter colors. We show in black, the star formation density for our different mass-metallicity models (lower legend).}
    \label{fig: eta_vs_Z}
\end{figure*}

Fig. \ref{fig: R0_BPSxCalib} depicts the values of the present-day astrophysical merger rate, $R_\mathrm{0} \equiv R_\mathrm{mBBH}(z_\mathrm{mBBH}=0)$, for our 240 models. For comparison, we show the 90\% confidence range of $R_\mathrm{0}$ inferred by the LVK collaboration (16 to 61 mergers $\mathrm{Gpc^{-3}yr^{-1}}$)  after the O3 observing run \citep{GWTC-3}. Our results have higher rates than the LVK and we discuss this comparison in more detail in \S \ref{sec: Discussion}. The merger rates span about 2 orders of magnitude. For a given BPS model, the rates typically span a factor of 4 across SFR/MZR models. Globally, BPS models do not show much variation in the merger rates, with a few outliers with large deviations. For illustrative purposes in the following analysis, we chose a subset of BPS models, highlighted in green, as representative of the 40 BPS models. These models were chosen to illustrate typical variations around the default model.

To understand the influence of the SFR/MZR models and mBBH formation efficiency on the merger rate $R_\mathrm{0}$, we show the interplay between the mBBH formation efficiency and the star formation as a function of metallicity in Fig.~\ref{fig: eta_vs_Z}. Across BPS models, the mBBH formation efficiency sharply drops beyond low metallicities ( $\gtrsim$ 0.1 \Zsun). This is due to increased mass loss from stellar winds resulting in a lower mass BH and a comparatively larger impact of SN kick. In contrast, in all SFR/MZR models, the star formation drops towards low metallicity. As described in \S\ref{sec: astro progen rate}, the progenitor formation rate depends on the product of the SFR and $\eta_\mathrm{mBBH}$ (Eq.~\ref{eqn: PopAstro rate}). As such, the region of \Z\ $\mathrm{ \simeq 0.1-0.3}$ \Zsun\ dominates the production rate of merging black holes. In contrast, regions of very low metallicity (\Z\ $\mathrm{\lesssim 0.01}$ \Zsun) do not contribute significantly because of the low star formation rate and the regions of very high metallicity  (Z $\mathrm{\gtrsim}$ \Zsun) do not contribute because of their low $\eta_\mathrm{mBBH}$. This comforts our choice of the metallicity range we use in our simulations (see \S\ref{sec: SFR}).

 From Fig.~\ref{fig: eta_vs_Z}, we can expect the largest influence on the progenitor formation rate in intermediate metallicities, around 0.1 to 0.3 \Zsun. Models with large efficiency in intermediate metallicities will tend to have higher progenitor formation rate and as a result have higher merger rates. As an example, the $\alpha\mathrm{(0.2)}\lambda\mathrm{(1)}$ model has consistently lower efficiency than the $v_\mathrm{BH}\mathrm{(3)}$ model for all metallicities except between the crucial 0.1 to 0.3 \Zsun. As a result, the merger rate of $\alpha\mathrm{(0.2)}\lambda\mathrm{(1)}$ is higher than that of $v_\mathrm{BH}\mathrm{(3)}$. 

For most BPS models, Fig.~\ref{fig: R0_BPSxCalib} shows a consistent trend in the rates across SFR/MZR models: KK04\_Ma16 and KK04 produce the lowest rates while PP04 and PP04\_Ma16 produce the highest rates. T04 and M09 fall in between and with nearly equal rates. Again, this trend can be attributed to the metallicity dependence of the SFR/MZR models and $\eta_\mathrm{mBBH}$, as shown in Fig.~\ref{fig: eta_vs_Z}. As the efficiency of mBBH formation sharply drops off for \Z\ $\gtrsim$ 0.2 \Zsun, SFR/MZRs that favor lower \Z\ (like PP04 and PP04\_Ma16) will produce higher merger rates while SFR/MZRs that favour high \Z\ (KK04\_Ma16 and KK04) will result in low merger rates. The similar SFR/MZRs of T04 and M09 result in similar, intermediate merger rates. The high-redshift extrapolation from \citet{Chruslinska+19} results in higher merger rates than the \citet{Ma+2016} extension due to a faster decrease of the global metallicity at high redshifts. 

The consistently large rates of PP04 and PP04\_Ma16 in comparison with that of GW observations indicate that either the default BPS model is overly optimistic and overproduces mBBH by more than an order of magnitude or that the calibration associated with the MZRs of PP04 and PP04\_Ma16 \citep{PP04} is in tension with the observations.

\subsection{Astrophysical progenitors environments}
\label{sec: Results-AstroProgen}

\subsubsection{Global trends}
\label{sec: Progenitor: global trends}
% ASTRO RATES
\begin{figure*}%[H]
    \centering
    \includegraphics[width=\linewidth]{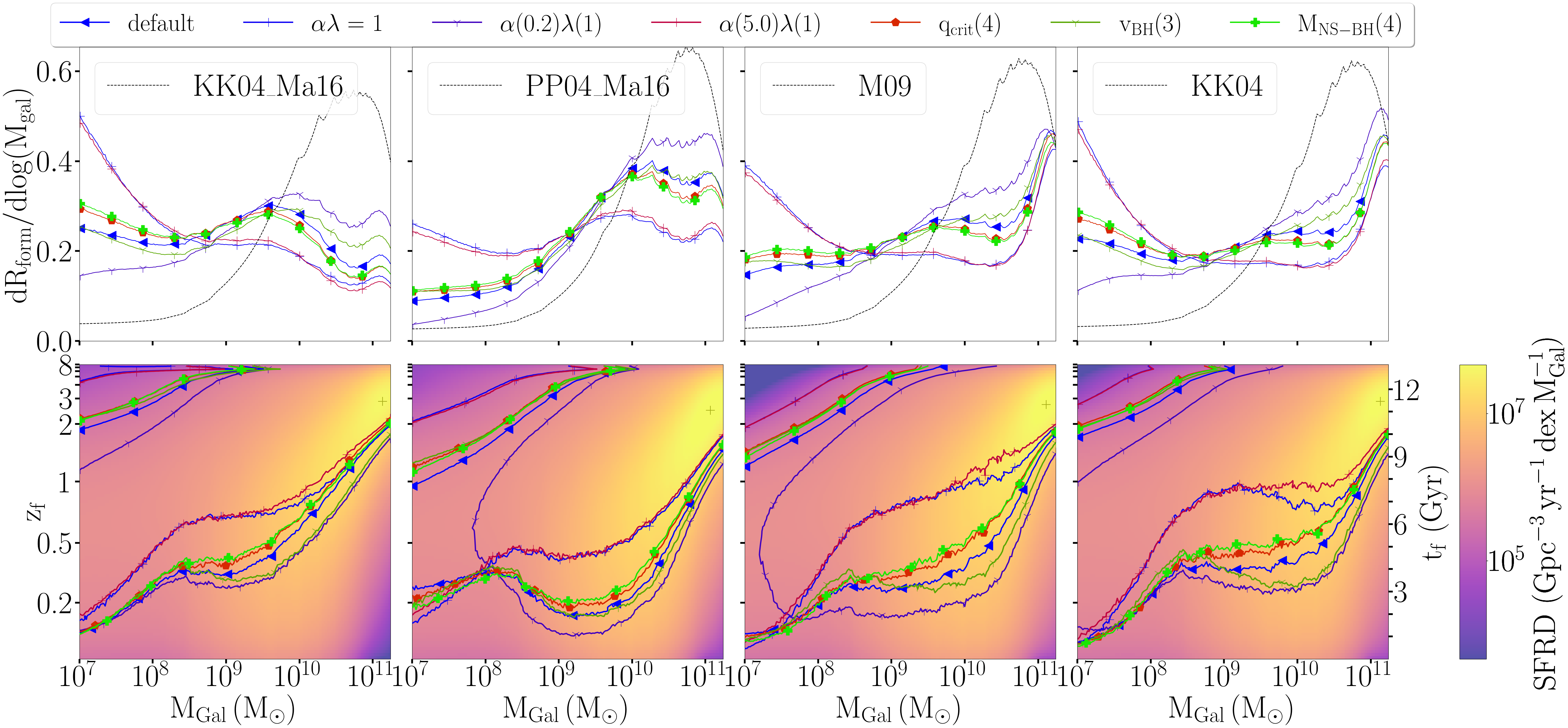}
   \caption{Typical formation galaxy mass and formation time of merging black holes. In the bottom row, we show the contours containing 90\% of the progenitor formation rate in terms of present-day galaxy mass and time of formation, for different binary population synthesis models (colored lines with symbols) and different mass-metallicity models (from left to right). The logarithmic color map depicts the global star formation rate in each model for comparison. In the top row, we show the same distributions, summed over formation time, with the corresponding star formation model shown with the black line.}

    \label{fig: AstroProgen_Mgal_z}
\end{figure*}

Fig. \ref{fig: AstroProgen_Mgal_z} depicts the progenitor galaxy mass and redshift (or equivalently look-back time) of the formation of mBBHs. 
The color map shows the global star formation rate density of each model.

In the 1-dimensional distributions (upper panels), all our models present a tri-modal distribution of the typical formation galaxy, with variations owing to the specific MZR or BPS model. We observe three peaks in progenitor formation rates at \Mgal\ = $\mathrm{10^7}$ \Msun, \Mgal\ $\simeq 3 \times 10^9$ \Msun and \Mgal\ $\simeq 10^{11}$ \Msun.  In comparison, the global SFRD is strongly dominated by massive galaxies (\Mgal\ $\gtrsim 2 \times 10^{10}$ \Msun). This strong SFRD at high \Mgal\ and high \zf\ ($\gtrsim$ 2) causes the corresponding high mBBH formation rate. As such, the peak of SFRD (cross in the bottom plots) always lies within the contour of mBBH progenitor formation rate. At high galaxy masses, the mBBH progenitor formation can be well mapped to the global star formation, with a small shift towards higher redshifts.

Away from the peak SFRD, at low redshifts (\zf $\mathrm{\lesssim}$ 1), the progenitor contours significantly deviate from the SFRD and favor dwarf galaxies (\Mgal\ < $\mathrm{10^9}$ \Msun) across most BPS models, irrespective of the SFR/MZR model. This is primarily due to the abundance of low metallicity star-forming gas in dwarf galaxies \S \ref{sec: SFR} which have the highest efficiencies of mBBH formation (Fig. \ref{fig: eta_vs_Z}). This effect causes the upturn for dwarf galaxies that is present in most of the models, except for the ones which have comparatively small efficiencies at low metallicity (\Z\ $\lesssim 0.1$ \Zsun), for e.g., the $\alpha\mathrm{(0.2)}\lambda\mathrm{(1)}$ model. In models with the highest efficiency at low metallicity ($\alpha\lambda$=1 and $\alpha\mathrm{(5)}\lambda\mathrm{(1)}$ models) formation in dwarf galaxies can be equal to (for the KK04 and M09 MZRs) or larger than (for the KK04\_Ma16 MZR) the contribution from the most massive galaxies. 

\begin{figure*}%[H]
    \centering
    \includegraphics[width=\linewidth]{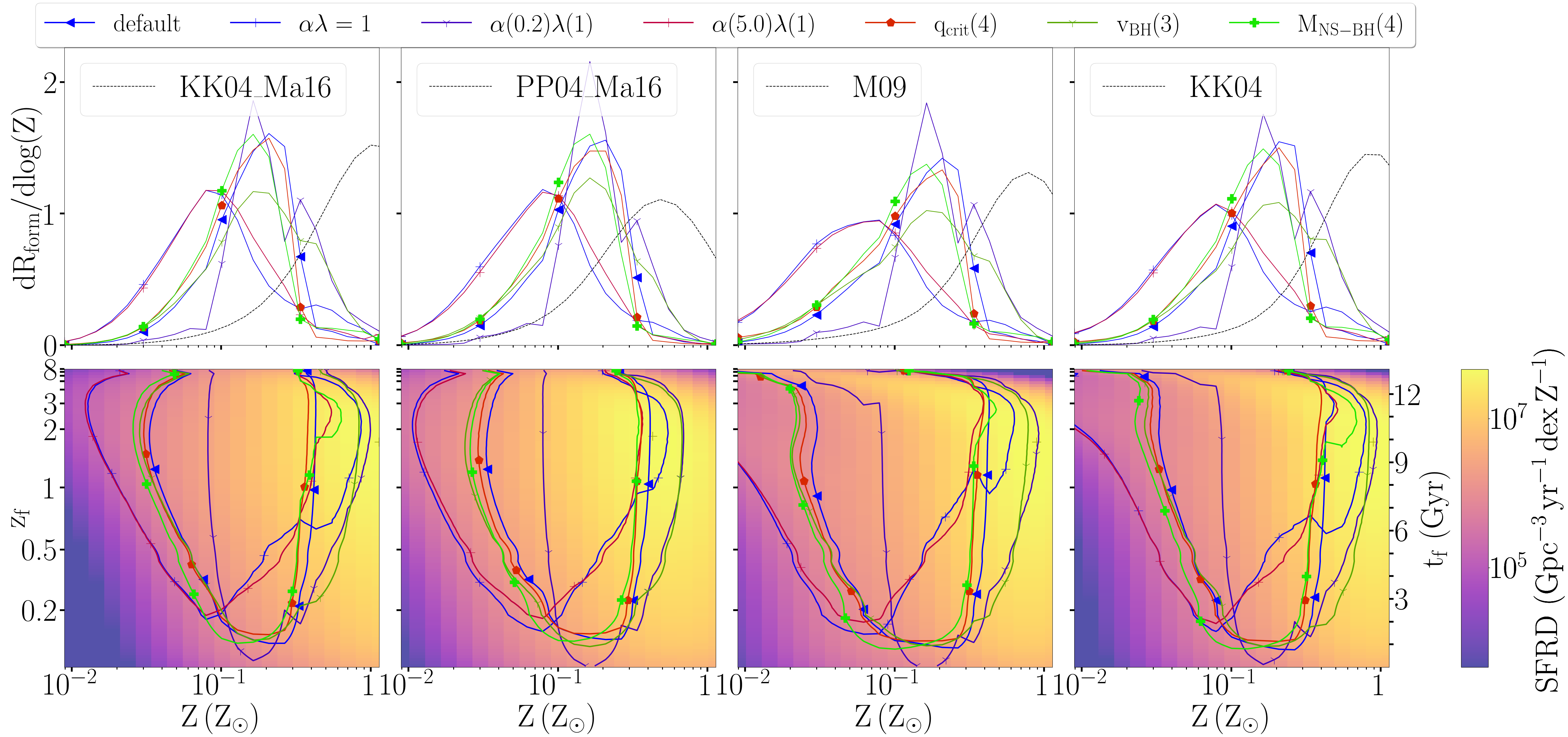}
    \caption{Typical formation galaxy metallicity and formation time of merging black holes. In the bottom row, we show the contours containing 90\% of the progenitor formation rate in terms of galaxy metallicity and time of formation, for different binary population synthesis models (colored lines with symbols) and different mass-metallicity models in each panel. The logarithmic color map depicts the global star formation rate in each model for comparison.  In the top row, we show the  same distributions, summed over all formation time, with the corresponding star formation model shown with the black line.}
    \label{fig: AstroProgen_Z_z}
\end{figure*}

Fig. \ref{fig: AstroProgen_Z_z} shows the progenitor \Z\ and \zf\ for the same models. In all our BPS models, the metallicity region of \unit[0.1-0.3]{\Zsun} strongly contributes or even dominates the formation of mBBH progenitors, irrespective of formation time. As an example, looking at the 1-dimensional plots (top row) for the default, $\alpha\mathrm{(0.2)}\lambda\mathrm{(1)},\,q_\mathrm{crit}\mathrm{(4)},\,\mathrm{and}\,M_\mathrm{NS-BH}\mathrm{(4)}$ BPS models, the fraction of systems with \Z\ $\mathrm{\in}$ 0.1 - 0.3 \Zsun\ is at least 50\%. As described in \S\ref{sec: Results-AstroMergerRates}, this is due to the interplay between the SFR which favors high metallicities and $\eta_\mathrm{mBBH}$ (Fig. \ref{fig: eta_vs_Z}) which favors low metallicities. Outside the 0.1-0.3 \Zsun\ metallicity range, the overlap across different BPS models can be very limited due to the large dependence on metallicity in BPS modeling. 

\begin{figure*}%[H]
    \centering

    \includegraphics[width=\textwidth]{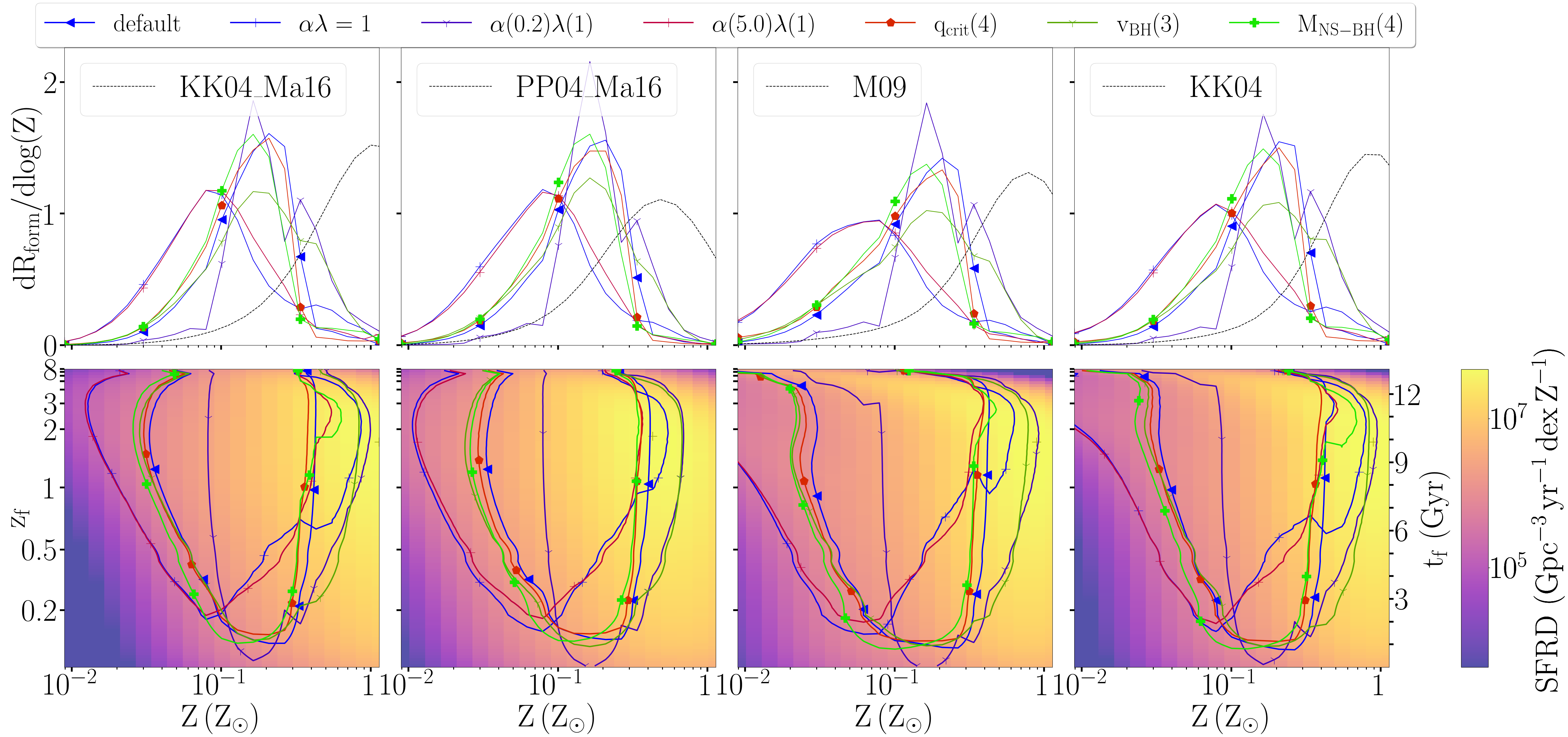}
    \caption{Typical formation galaxy mass of merging binary black holes. The logarithmic color map shows the \Mgal\ with the maximum progenitor formation rate for a given \Z\ and \zf. The colored contours contain 90\% of the progenitor formation rate of our default binary population synthesis model and the black contours contain 90\% of the star formation rate for different mass metallicity models. The white spaces correspond to the progenitor formation rates that fall below the resolution of our simulation.}
    \label{fig: AstroProgen3D_Z_z}
\end{figure*}

Looking at the 2-dimensional \zf-\Z\ plane of Fig. \ref{fig: AstroProgen_Z_z}, throughout  cosmic history, the progenitor metallicity shows a wide distribution centered at \unit[0.1-0.3]{\Zsun}.  For all models, the metallicity distribution of the progenitors narrows with decreasing redshift, and again, this can be attributed to the combination of the SFR and the efficiency of mBBH formation, $\eta_\mathrm{mBBH}$. The plots of M09 and KK04 calibrations show a correlation at high redshift (\zf\ $\mathrm{\gtrsim \, 3}$) between the contour of the progenitor formation rate and the global SFR. This is the region of the SFR which is extrapolated based on the assumption of the high redshift evolution of the MZR. In these SFR/MZR models,  most of the star formation occurs at low metallicity (\Z\ $<$ \unit[0.1]{\Zsun}) for $z_f>3$ (see the yellow extension towards the upper left corner in Fig.~\ref{fig: SFR_Z_z}). Fig. \ref{fig: eta_vs_Z} shows that the efficiency of mBBH formation does not vary significantly in this metallicity region. Therefore, the BPS contours directly trace the SFR at high \zf.
With decreasing \zf\, the increasingly high metallicity causes the lower metallicity limit of the colored contours containing 90\% of the progenitor formation rate to shift slightly towards higher metallicities. 
However, a deciding factor that places an upper limit on the metallicity of progenitors (\Z\ $\mathrm{\lesssim}$ 0.5 \Zsun) is the nearly constant $\eta_\mathrm{mBBH}$ for low metallicities (\Z\ $<$ \unit[0.1]{\Zsun}) and its rapid decline for higher values (\Z\ $>$ \unit[0.2]{\Zsun}). Thus the progenitor contours generally favor lower metallicities as compared to the SFR, while also slightly shifting towards higher metallicities at lower redshifts, hence its narrowing shape. 

To help elaborate further, Fig. \ref{fig: AstroProgen3D_Z_z} shows the mode of the distribution of formation galaxy masses of mBBH progenitors for a given \Z\ and \zf\ for the default BPS model. Low metallicity progenitors tend to come from dwarf galaxies (\Mgal\ $<$\unit[$10^9$]{\Msun}) while larger galaxies have higher metallicities. At large \zf, most of the progenitor formation is driven by the SFR, meaning that across \Z, the progenitors are mostly present in galaxies that would now be massive. At lower \zf, stars in massive galaxies have increasingly high metallicity while low metallicity environments can be found in dwarf galaxies.

Looking closely at the trends in individual BPS models, the $\alpha\mathrm{(0.2)}\lambda\mathrm{(1)}$ model stands out along the three dimensions of our star formation models (see Figs.~\ref{fig: AstroProgen_Mgal_z}-\ref{fig: AstroProgen_Z_z}). The  $\eta_\mathrm{mBBH}$ of this specific BPS model has a peculiar shape with low efficiency at high and low \Z\ while peaking between 0.1 and 0.2 \Zsun. As a result, the progenitor formation rate is concentrated within 0.1-0.2 \Zsun. This implies that the formation rate contours are also smaller in \Mgal\ vs \zf\ and \Z\ vs \zf. On the other hand, the contours for $\alpha\lambda\mathrm{(1)}$ and $\alpha\mathrm{(5.0)}\lambda\mathrm{(1)}$ models show a large overlap as a result of their very similar $\eta_\mathrm{mBBH}$(\Z) seen in Fig. \ref{fig: eta_vs_Z}. 

\subsubsection{Effects of the black hole mass}

% M1 BBH Plots
\begin{figure}%[H]
    \centering
    \includegraphics[width=.49\textwidth]{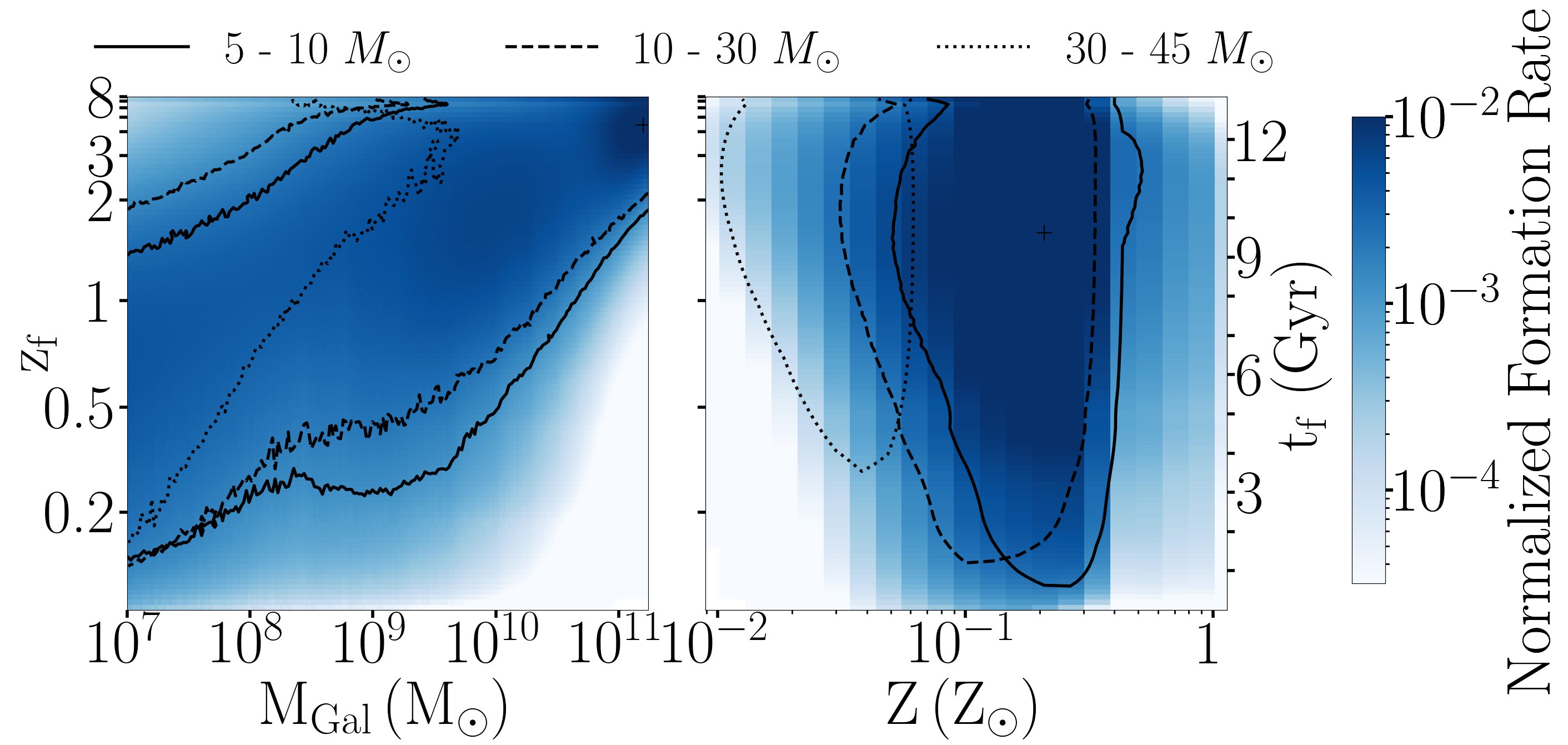}
    
    \caption{Typical progenitor galaxy mass, metallicity, and time of formation for merging black holes with different ranges of primary black hole mass, $M_\mathrm{1, BBH}$. The logarithmic color map shows the normalized progenitor formation rate for all mass ranges. The contours contain 90\% of the progenitor formation rate, for different $M_\mathrm{1, BBH}$ ranges (legend) as a function of \Mgal\ vs \zf\ (left) and \Z\ vs \zf\ (right). We use our default model of the mass-metallicity relation (KK04\_Ma16) and binary population synthesis.}
    \label{fig: AstroProgen_M1BBH}
\end{figure}

Taking our default SFR/MZR model and BPS model (KK04\_Ma16, default respectively), we look at the typical progenitor environment for black holes of different masses. For simplicity, we consider the mass of the primary black hole, defined as the larger of the two merging black holes. In practice, almost all our BHs have an almost equal mass ratio. 

Fig. \ref{fig: AstroProgen_M1BBH} shows the contours containing 90 \% of the formation of BH merger progenitors for three ranges in primary black hole mass: 5-10 \Msun, 10-30 \Msun, and 30-45 \Msun. The color map shows the overall normalized progenitor formation rate. The plot on the right shows that with increasing primary black hole mass, the progenitor contours favor lower metallicities (see ~\S\ref{sec: BPS}). The black hole mass only indirectly depends on \Mgal\ and \zf\ which correlates with \Z\ through the SFR. For example, large black holes tend to be formed from low-metallicity progenitors. We have shown in Fig. \ref{fig: AstroProgen3D_Z_z} that low metallicity progenitors are found both at low redshifts in typically low-mass galaxies and at high redshifts in a wide range of galaxy mass. Combining all BH masses (color map), we recreate the global trends in the progenitor formation rate that is shown in Figs \ref{fig: AstroProgen_Mgal_z} and \ref{fig: AstroProgen_Z_z}.

\subsection{Progenitors environments of detected merging black holes}
\label{sec: Results-DetProgen}
% DETECTED RATES
\begin{figure*}%[H]
    \centering
    \includegraphics[width=\linewidth]{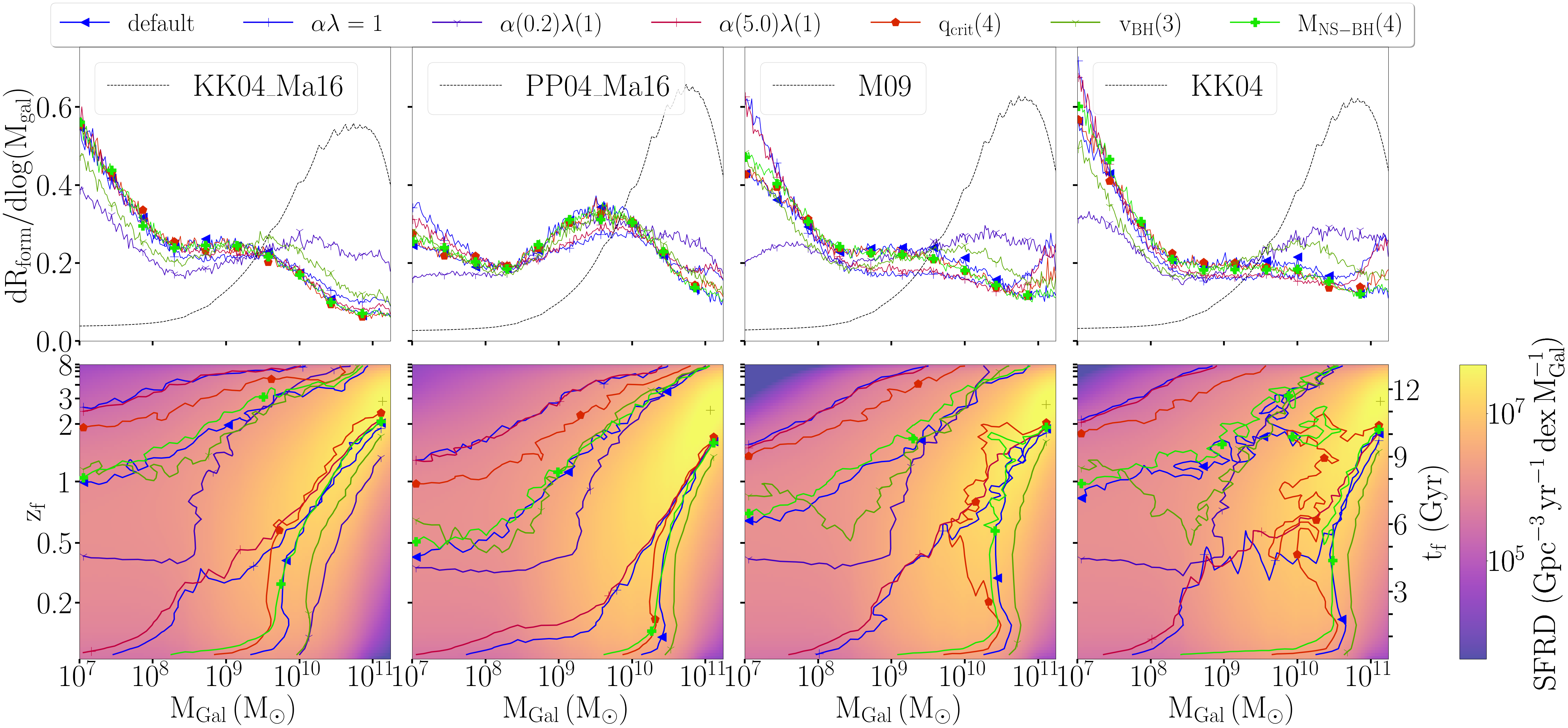}
    \caption{Typical formation galaxy mass and formation time of detectable merging black holes. In the bottom row, we show the contours containing 90\% of the progenitor formation rate in terms of galaxy mass and time of formation, for different binary population synthesis models (colored lines with symbols) and different mass-metallicity models in each panel. The logarithmic color map depicts the global star formation rate in each model for comparison. In the top row, we show the same distributions, summed over all formation time, with the corresponding star formation model shown with the black line.}
    \label{fig: DetProgen_Mgal_z}
\end{figure*}

Using the astrophysical population of mBBH, we explore the detected population and the effect of GW detector selection effect on the inference of the progenitor formation environment. Fig. \ref{fig: DetProgen_Mgal_z} shows the progenitor \Mgal\ and \zf\ for mBBHs that are detectable by the LIGO-Virgo three-detector network in the O3 configuration. The 1-dimensional plot (top) shows fewer variations across BPS models as compared to the astrophysical progenitors in Fig. \ref{fig: AstroProgen_Mgal_z}. This indicates that the bias introduced by current detectors largely limits the fraction of the astrophysical mBBH population they can observe.

\begin{figure*}%[H]
    \centering
    \includegraphics[width=\linewidth]{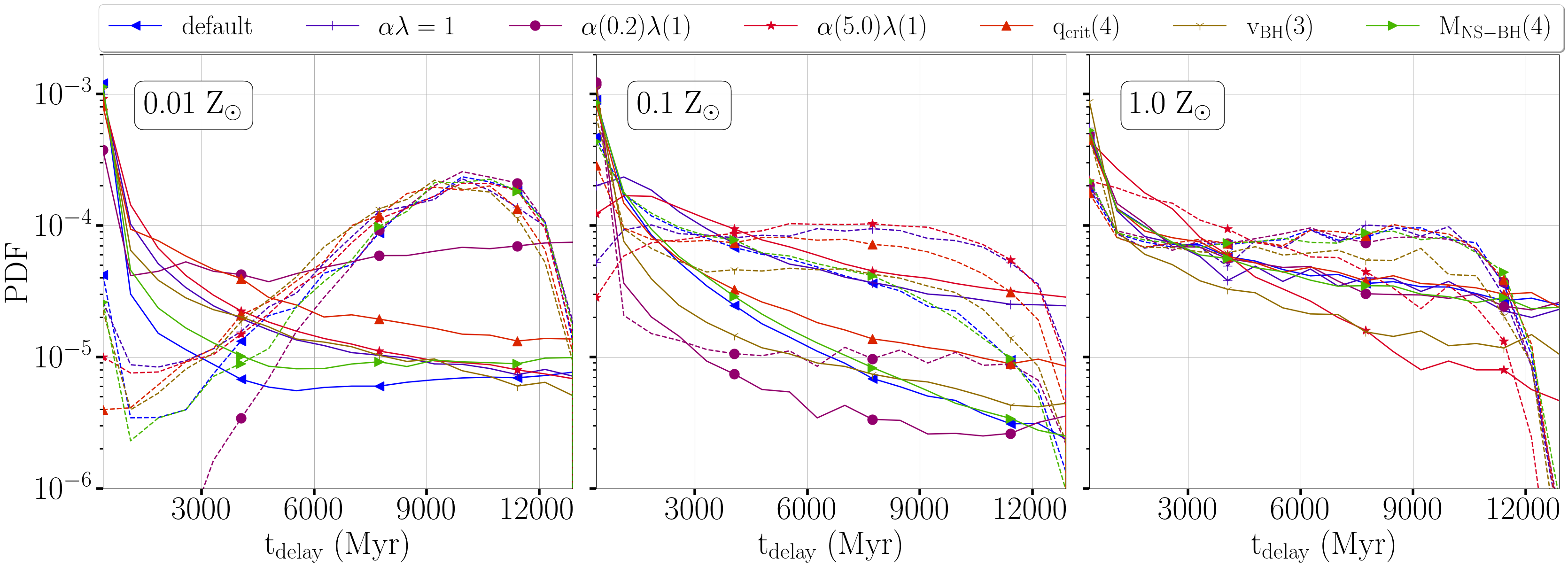}
    \caption{Normalized histogram of the delay time (\tdelay) of the mBBHs from the \PopBPS\ of the binary population synthesis simulation (solid lines), and from the subset of the astrophysical population \PopAstro\ that are detectable, (dashed lines) for different BPS models (colors and symbols), across 3 metallicity bins: 0.01 \Zsun (left), 0.1 \Zsun (center), 1 \Zsun (right).}
    \label{fig: t_delay_BPS}
\end{figure*}

\begin{figure*}%[H]
    \centering
    \includegraphics[width=\linewidth]{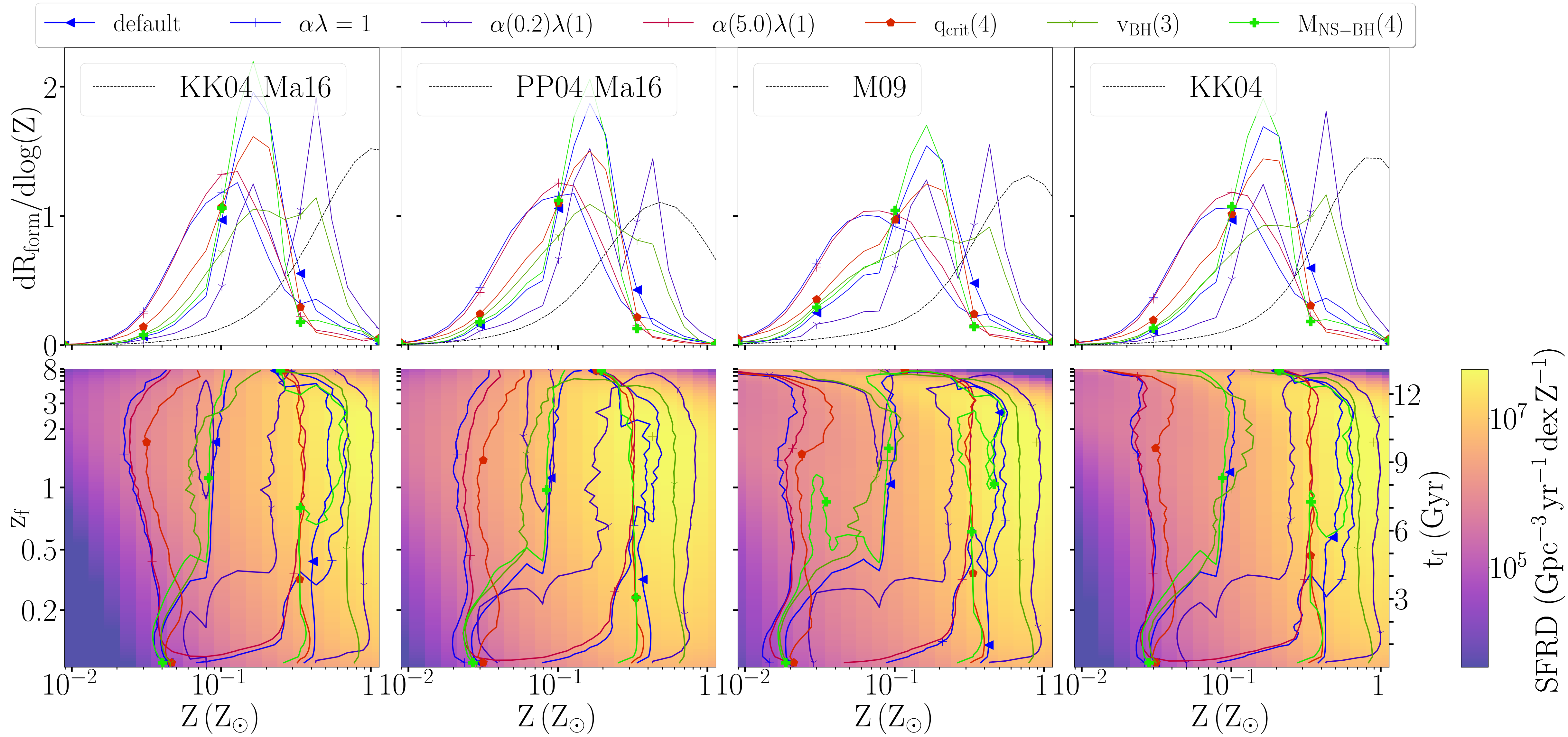}
    \caption{Typical formation galaxy metallicity and formation time of detectable merging black holes. In the bottom row, we show the contours containing 90\% of the progenitor formation rate in terms of galaxy metallicity and time of formation, for different binary population synthesis models (colored lines with symbols) and different mass-metallicity models in each panel. The logarithmic color map shows the global star formation rate in each model for comparison.  In the top row, we show the  same distributions, summed over all formation time, with the corresponding star formation model shown with the black line.}
    \label{fig: DetProgen_Z_z}
\end{figure*}

\begin{figure}%[H]
    \centering
    \includegraphics[width=.49\textwidth]{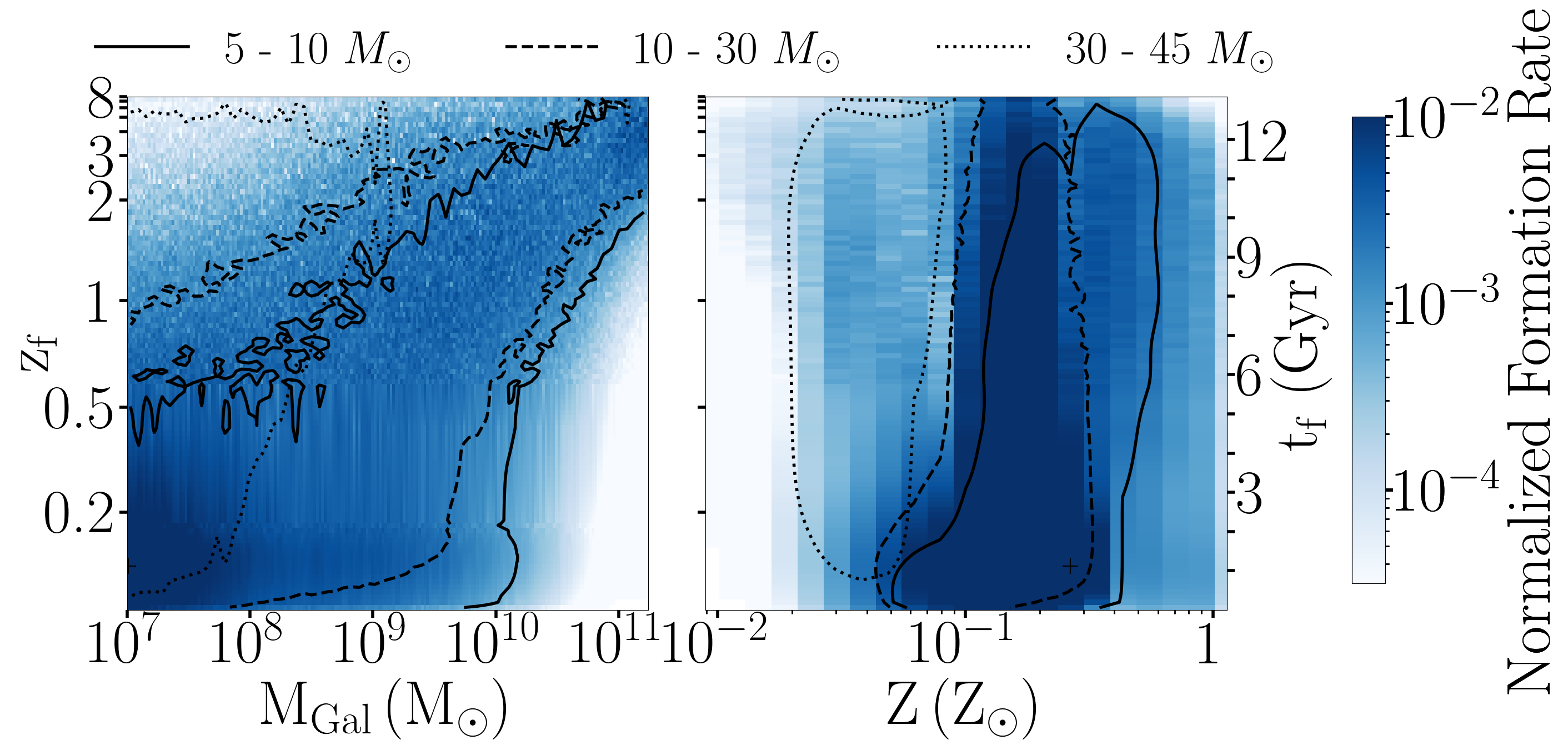}
    
    \caption{Typical progenitor galaxy mass, metallicity, and time of formation for \textit{detected} merging black holes with different ranges of primary black hole mass, $M_\mathrm{1, BBH}$. The logarithmic color map shows the normalized progenitor formation rate for all mass ranges. Contours contain 90\% of the progenitor formation rate, for different $M_\mathrm{1, BBH}$ ranges (legend) as a function of \Mgal\ vs \zf\ (left) and \Z\ vs \zf\ (right). We use our default model of the mass-metallicity relation (KK04\_Ma16) and binary population synthesis.}
    \label{fig: DetProgen_M1BBH}
\end{figure}

Moreover, there is a large overlap in the contours of the 2-dimensional plot (bottom) in two distinct progenitor environments: high \zf\ (> 2), large \Mgal\ (>$\mathrm{10^{10.5}}$ \Msun) progenitors and low \zf, small \Mgal\ (< $\mathrm{10^{10}}$ \Msun) progenitors. Comparing with the astrophysical progenitors in Fig. \ref{fig: AstroProgen_Mgal_z}, it is evident that the abundance of astrophysical progenitors at high \zf, in large \Mgal\ has influenced the progenitors of detected mBBHs in the same parameter space, albeit, with a very low formation rate as shown in the 1-dimensional plot (KK04\_Ma16 with default BPS progenitor formation rate in massive galaxy fall by a factor of 2 in the detected population).

Across all models, the progenitor formation at low  \zf\ in small \Mgal\ (< $\mathrm{10^{10}}$ \Msun) galaxies is more prominent than in the case of the astrophysical population. The inverse relation between the signal-to-noise ratio (SNR) and luminosity distance implies that the detectors are biased towards a low redshift of the merger. To analyze the relationship between the redshift of merger and the redshift of progenitor formation, we look at the distribution of the delay time defined as the time interval between the formation of the binary stars to the merger of BBH. Fig. \ref{fig: t_delay_BPS} shows the delay time distribution of mBBHs for 3 metallicities for systems from the BPS simulation \PopBPS\ (solid lines), and for the subset of the astrophysical population \PopAstro that are detectable (dashed lines). We do not show the global \PopAstro\ as they closely follow the BPS curves, but with a sharp drop in the number of systems with \tdelay\ $\mathrm{\gtrsim}$ 10 Gyr as these systems generally fail to merge by the present-day. The vast majority of delay times in the BPS simulation are short, implying that the merger redshift closely follows the formation redshift \zf\ for the astrophysical population. However, considering the detectable mBBH population, the bias towards detecting mergers at low redshift, in combination with the influence of the SFR towards high \zf\ progenitors, results in preferred long delay times, especially at lower metallicities.

Fig. \ref{fig: t_delay_BPS} shows that large delay times are common for higher metallicity progenitors. As such, we expect high redshift progenitors of detected systems to have \Z\ > 0.1 \Zsun. Fig. \ref{fig: DetProgen_Z_z} indeed shows that the concentration of progenitors at high \zf\ corresponds to higher \Z\ (> 0.1 \Zsun). This also contributes to the slight shift towards higher metallicities in the 1-dimensional plot (upper row) as compared to the astrophysical metallicity distribution (Fig. \ref{fig: AstroProgen_Z_z}). The first peak in the 1-dimensional plot between 0.1-0.3 \Zsun\ is due to the astrophysical bias, while the second peak around \Z\ $\simeq$ 0.5 \Zsun\ due to the large delay times high \zf\ mBBHs. Furthermore, there exists a secondary detector bias towards larger black holes due to their larger GW signal amplitude which plays a role in selecting low metallicity progenitors. This can be seen in the low \zf\ progenitors of the 2-dimensional plots of Fig. \ref{fig: DetProgen_Z_z}. This effect is not quite evident in higher \zf\ as large delay times favor large \Z. 

Fig. \ref{fig: DetProgen_M1BBH} shows the progenitor distribution of detected mBBHs for different primary black hole masses. The majority of detected mBBHs come from dwarf galaxies, at low \zf, and with metallicity of 0.1 to 0.3 \Zsun. This is in good agreement with existing progenitor predictions of GW events \citep{Lamberts2016}. Large black holes are predominantly formed in low mass galaxies (< $\mathrm{10^{10}}$ \Msun) and with lower metallicities ($\mathrm{\leq}$ 0.1 \Zsun) than smaller black holes which arise from a wide range of galaxy masses and higher metallicities ($\mathrm{\geq}$ 0.3 \Zsun).

\subsection{Progenitor galaxy posterior of GW150914}
\label{sec: Results-ProgenPosterior}
As a follow-up of our simulations, we infer the properties of the progenitor formation galaxy of detected mBBHs in LIGO-Virgo data. To produce the posterior probability distribution of \Mgal, \Z, and \zf\ for any detected GW event, we compute the probability of the progenitor formation rate $p_{\rm sim}$ as a function of the component black hole masses $M_\mathrm{{1, BBH}}$, $M_\mathrm{{1, BBH}}$ and luminosity distance $D_\mathrm{Lum}$, weighted by the corresponding detected posterior distribution of the GW event as shown below

\begin{eqnarray}
p(M_\mathrm{Gal}, Z, z_\mathrm{f} | x_{\rm GW}) = \int p(M_\mathrm{{1, BBH}}, M_\mathrm{{2, BBH}}, D_\mathrm{Lum} | x_{\rm GW}) \nonumber \\ 
\times \,\, p_{\rm sim}(M_\mathrm{Gal}, Z, z_\mathrm{f} | M_\mathrm{1, BBH}, M_\mathrm{2, BBH}, D_\mathrm{Lum}) \nonumber \\ 
\times \,\, \mathrm{d}M_\mathrm{1, BBH} \, \mathrm{d}M_\mathrm{2, BBH}  \, \mathrm{d}D_\mathrm{Lum},
\end{eqnarray}
where $x_{\rm GW}$ represents the GW event. This simple statistical method involves re-weighting the typical \Mgal, \Z\ values from the simulation by the most credible mass and redshift values of detected GW events. In the hypothesis that a given detected GW is formed by the processes described by the simulation, the method rapidly provides its most credible values for \Mgal, \Z. In the case that the formation channel is not known, or the detection rate and GW event are not consistent with the simulation, one should use a hierarchical inference approach (e.g. \citealp{Bouffanais_21_inference_channel,Simone_InferStellarProgen}) considering other formation channels. Here, we are not interested in considering multiple formation channels but instead provide the most credible \Mgal\ and \Z\ assuming that a particular GW event can be described with our simulation.

\begin{figure*}
    \includegraphics[width=\textwidth]{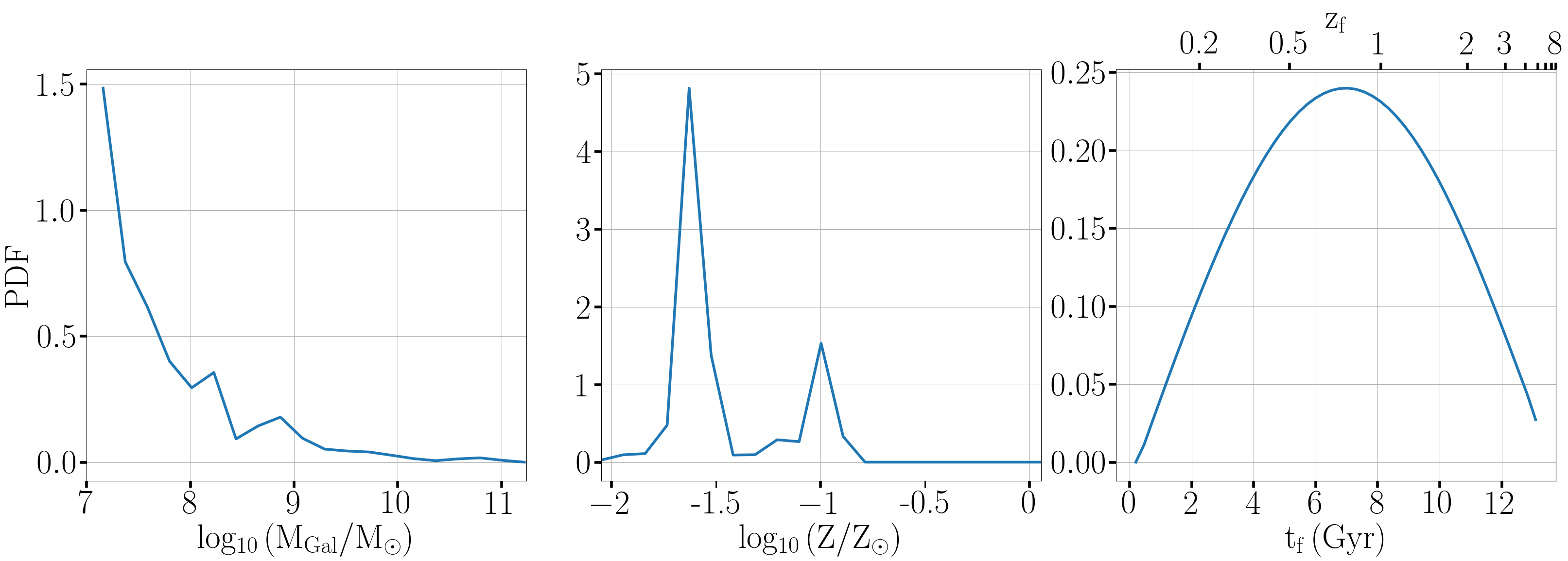}
    \caption{Posterior probability density function (PDF) of the progenitor galaxy mass \Mgal, metallicity \Z\ and time of formation \tf\ for the first detected gravitational wave event (GW150914) based on our default star formation model (KK04\_Ma16) and binary evolution model (default COSMIC). This illustrates our method to produce progenitor posteriors given posterior samples of a detected merger.}
    \label{fig: ProgenitorPosterior}
\end{figure*}

As an illustration, Fig. \ref{fig: ProgenitorPosterior} shows the progenitor environment of the first event, GW150914 \citep{GW150914}, for our default model (KK04\_Ma16 with default BPS flags). It distinctly shows that GW150914 likely came from a dwarf galaxy (\Mgal $\mathrm{\lesssim \, 10^9}$\Msun) with very low metallicity (\Z $\mathrm{\simeq}$ 0.025 \Zsun), and with a broad distribution of progenitor formation look-back time which peaks at a corresponding redshift of \zf\ $\mathrm{\simeq}$ 1. This is consistent with the initial study by \citep{Lamberts2016} of the GW event. The secondary peak in metallicity follows from the peak in astrophysical formation rates at around 0.1 \Zsun, as discussed in detail in \S \ref{sec: Progenitor: global trends}. The primary peak in metallicity is due to GW150914's combination of a high-mass BBH merging at close proximity. We can see in Fig. \ref{fig: BPS}  that a remnant mass BH $\mathrm{\simeq}$ 70-80 \Msun, which was the mean total mass of GW150914, can be formed from progenitors with \Z $\mathrm{\simeq}$ 0.01 \Zsun\ (consistent with Figure \ref{fig: DetProgen_M1BBH}). This low metallicity is typically seen in dwarf galaxies, especially in low \zf, as shown in Fig. \ref{fig: AstroProgen3D_Z_z}. The broad posterior distribution in \tf\ is due to the presence of dwarf galaxies with low metallicities at almost all redshifts, as can be inferred from Fig. \ref{fig: AstroProgen3D_Z_z}. We find that the progenitor posteriors prediction from other BPS and SFR/MZR models broadly agree that GW150914 had a low metallicity (\Z\ $\lesssim\,0.1$\Zsun) dwarf galaxy (\Mgal\ $\lesssim\,10^9$ \Mgal) progenitor with small deviations depending on the different values of $\eta_{\mathrm{mBBH}}$(\Z) value of each BPS model.

\section{Discussion}
 \label{sec: Discussion}

Our 240 combinations of SFR/MZR models and binary evolution models result in a wide range of BBH merger rates and inferred formation galaxy properties. This globally suggests that caution should be used when interpreting BBH merger simulations. Still, global trends arise from our analysis. The uncertainties of the mass-metallicity relations (MZR) result in rate differences of roughly a factor 4, which is higher than the uncertainties of the bulk of the binary models we have explored here. This is consistent with other works, such as \citet{Neijssel+2020, Tang2020_BBH_vs_MZR, Briel_22_transient_rates}, which find that differences in the star formation model, which includes the global star formation rate, mass metallicity relation and its spread, and galaxy stellar mass function, dominate the differences between the binary evolution models they explore, both for models based on simulations and observations. \citet{Broekgaarden_21_BBH_MZR_vs_bin, Santoliquido_21_SFR_vs_binary} confirm these findings in a systematic exploration of models.

This suggests that the choice of MZR, and particularly its high redshift extension is a key parameter in modeling BBH mergers. Based on the models we have explored, MZRs with more low-metallicity star-forming gas such as the one from \citet{PP04} are difficult to reconcile with observed BBH rates, especially when the low metallicity evolution is extrapolated beyond \zf\ $\mathrm{\simeq 3.5}$ \citep[as in][]{Chruslinska+19}. Fitting these MZR models with the measured LVK rates would require a typical decrease in the formation of merging BBH of more than an order of magnitude. The majority of our BPS models would require a reduction of a factor 2.

Our star formation model also includes a global star formation rate based on \citet{behroozi2013} and a galaxy stellar mass function based on \citet{Tomczak+2014}. As was detailed in \citet{Chruslinska+19}, our galaxy stellar mass function includes more low-mass galaxies than other models. Additionally, \citet{Applebaum_stochasticIMF_DwarfGalaxies} show that the SFR in ultra-faint dwarf galaxies is suppressed. As we show in Fig.~\ref{fig: AstroProgen_Mgal_z}, these dwarf galaxies can be large contributors to the BBH merger rates. As such, the uncertainties on the number of low-mass galaxies, at all redshifts, are another important driver of uncertainties. \citet{Neijssel+2020} showed that different galaxy mass functions can lead to a change in detected BBH merger rates up to a factor of 5.

Additional uncertainties stem from  the exact  initial conditions of our BPS models, such as the binary fraction and the initial masses and orbital parameters. The binary fraction of the most massive stars seems to be independent of metallicity \citep{Moe17_binaryProps}. \citet{Moe17_binaryProps} found correlations between the stellar masses, periods, and eccentricities, which we have ignored here, instead, using the data from \citet{sana12}. \citet{Klencki_18_newbin_paramsMoe} find that with these updated parameters, the global rates decrease by about a factor 2, while \citet{Tang2020_BBH_vs_MZR} show an increase of factor two for their binary model. These different findings indicate, once again, that caution should be taken when introducing new uncertainties and that generalization should be done with caution. 

Almost all our models predict effectively too high merger rates in comparison to observations (Fig. \ref{fig: R0_BPSxCalib}) with the values of $R_\mathrm{0}$ spanning almost 2 orders of magnitude. It is important to note that we have only chosen to explore 1-dimensional variations around the default choices (see Table \ref{tab:BPS flag} of Appendix \S \ref{sec: Appendix}). Therefore, we can conclude that the consistently large rates irrespective of the SFR/MZR model show evidence that the underlying default BPS model parameters proposed by \textsc{COSMIC} are too optimistic in forming mBBHs and are not the ones governing BBH formation in the Universe. As this study focuses on progenitor environments, we choose not to tune the BPS models in order to match the observed rate.  However, we expect that certain combinations of parameters can potentially lead to rates that are compatible with the observation. As an illustration, implementing the KK04\_Ma16 SFR and the default BPS model with two variations: $\alpha\mathrm{(0.5)}\lambda\mathrm{(1)}$ and $q_\mathrm{crit}\mathrm{(4)}$ results in a merger rate $R_\mathrm{0}$ = 25 $\mathrm{Gpc^{-3}\,yr^{-1}}$, within the boundaries of the observed rate.

It is important to note that BPS modeling is an ongoing field of research, with large uncertainties and oversimplification of complex mechanisms like the common envelope in unstable mass transfers and supernova kicks. We are severely limited by sparse observations due to the dynamic and transient nature of these mechanisms and defining a preferred model remains a challenge. 

Combined uncertainties on both binary/stellar evolution and star formation make the inference of the birth conditions (formation galaxy mass, metallicity, and cosmic time) non-trivial, especially for the typical formation galaxy mass. Still, we find that in all the models we explored, the most likely progenitor metallicity is around Z$\simeq$ 0.2 \Zsun and is narrowly peaked. \citet{CPBerry_Link_HMXRB_GW} have also found similar trends in the mBBH progenitor metallicity, further highlighting its significance. This effect translates from the interplay between the mBBH formation efficiency, and the star formation rate as a function of metallicity (Fig.~\ref{fig: eta_vs_Z}). Based on the range of models presented here, we find that one can predict the typical progenitor metallicity for merging BBHs formed through different BPS models provided the corresponding $\eta_\mathrm{mBBH}\mathrm{(Z)}$ and a star formation model.

The determination of the typical formation times is also rather consistent across models, with a steady increase of progenitor formation up to a look-back time of $\simeq$ 10 Gyr. Beyond that, we find that the formation rate decreases in our models using the \citet{Ma+2016} metallicity evolution at high redshift. On the other hand, models based on the \citet{Chruslinska+19} extrapolation at high redshift show a sharp peak of progenitor formation more than 12 Gyr ago, owing to the low metallicity gas. This is consistent with the findings of \citet{Graziani_20_BBH_Gamesh}, who find a typical formation time lies between 6 < \zf < 8. Within the models we have explored, different binary evolution models show limited differences in their typical formation times. This is because the delay time distributions prefer short timescales and the combination of the wide range of metallicities tends to wash out strong features in the delay time distributions.

The largest uncertainty is for the determination of the typical formation galaxy. Fig.~\ref{fig: AstroProgen_Mgal_z} shows three peaks whose heights vary strongly depending on both binary and MZR models. Extracting global trends of the progenitor environment based on the mBBH formation efficiency and global star formation model seems unclear at this point. Still, our results show agreement with the analysis of \citet{Santoliquido_22_host_galaxies} where the authors simulate the merger rate, formation galaxy and merger host galaxy of compact objects using observational scaling relations. Using an MZR similar to ours \citep{Chruslinska+19} to build their SFR, and BPS models probing different values of the CE efficiency ($\alpha$), they find that mBBHs tend to form in low mass, low metallicity galaxies, with a metallicity mode $\mathrm{\simeq\,0.1}$ \Zsun for a range of redshifts. They additionally show that BBHs tend to merge in more massive galaxies, as dwarf galaxies merge into more massive ones.

In this work, we have only considered the formation of mBBH through binary evolution, while dynamical environments are likely significantly contributing to the observed mergers as well \citep{Wong_21_channel_inference,Bouffanais_21_inference_channel,Zevin_21_channel_inference}. The uncertainties on typical conditions of formation of clusters \citep[e.g.][]{Brodie_Strader_06_cluster_review} and the uncertainty on the associated merger rates add an additional layer of ambiguity to the inference of the typical formation conditions of mBBHs.  

Ultimately breaking the degeneracies between initial conditions (global star formation rate and metallicity evolution) and stellar and binary evolution will require including additional information. First, one can limit the sample of explored models by only including those that have merger rates compatible with measurements. Then one can take this a few steps further by also including the information on the mass and possibly spin distributions. All of these will ultimately be measured as a function of redshift as detectors improve. With third-generation detectors, the astrophysical and detected distributions will be very similar,  removing the effects of detector biases from the analysis. Along the same lines, improved measurements of metallicity, star formation rates, and the galaxy stellar mass function, especially at high redshift will progressively decrease the allowed range for star formation models.

Still, degeneracies are likely to remain between star formation models and binary stellar evolution models. Joint modeling and observations of other types of transients associated with massive stars will be key. Neutron star mergers, and neutron star-black hole mergers to some extent, are much less dependent on stellar metallicity, and their merger rate more directly traces the global star formation rate \citep{Mapelli+2018hostgalaxiesDCO, Artale+2020_hostgalaxy, Santoliquido_22_host_galaxies}.

As such, most of the differences arise due to the different assumptions on binary and stellar evolution. \citet{Eldridge+2018_GWEMrates} and \citet{Briel_22_transient_rates} showed population models of an ensemble of transients, including core-collapse supernovae, long gamma-ray bursts, and pair-instability supernovae, which are likely related to the same underlying population of massive stars. They show how transients of different delay times are sensitive to different uncertainties. Combined modeling of a range of phenomena seems to be the way forward when breaking degeneracies and ultimately determining likely progenitor environments. 

Future research could extend the analysis to include massive black hole mergers as those formed for instance in active galactic nuclei and binary neutron star mergers. These sources have potential electromagnetic counterparts which can be used to test our estimation of the galaxy posterior. Such an analysis would entail BPS simulations of neutron star mergers and the simulation of dynamical black hole mergers in the dense environment of active galactic nuclei.

\section*{Conclusion}
To understand the progenitor environments of merging binary black holes (mBBHs), we have created 240 models, built using 6 mass-metallicity relations in the star formation rate (SFR) and 40 binary population synthesis (BPS) models. We find that the default BPS model and the variations explored in our analysis are overestimating the merger rate when compared to GW observations by about a factor of three for the KK04\_Ma16 model and by an order of magnitude for the PP04\_Ma16 model. Globally, the merger rates of mBBHs do not show significant variations across BPS models, although with a few outliers with large deviations. In contrast, the choice of the SFR can influence the rate by up to a factor of 4.  Globally, the merger rate for our different models results from the interplay between the SFR and the efficiency of the formation of mBBHs as a function of metallicity. BPS models with large efficiencies at intermediate metallicities (0.1 to 0.3 \Zsun) have large progenitor formation rates which contribute to large merger rates.

SFR/MZR models that predict high formation rates of low metallicity stars lead to high BBH merger rates. As a result, the redshift extrapolation of the \citet{Chruslinska+19} model has consistently higher merger rates than that of \citet{Ma+2016}. Moreover, the large BBH merger rates predicted by the  \citet{PP04} mass-metallicity calibration in comparison with GW observations indicate that either all BPS models are overly optimistic or that \citet{PP04} calibration is in tension with GW observations.

When trying to infer to typical progenitor environment of merging black holes, we find that most systems come from regions with
\Z\ $\mathrm{\simeq 0.1\,-\,0.3}$ \Zsun, regardless of the model. On the other hand, we find strong variations in the typical formation galaxy mass, depending both on the metallicity and the binary evolution model.

Nevertheless, the inferred formation galaxy distribution typically presents three peaks of varying importance: present-day dwarf galaxies with low metallicity (\Mgal\ $\mathrm{\lesssim 10^{8}}$ \Msun, \Z\ $\mathrm{\lesssim0.05}$ \Zsun), larger galaxies with intermediate metallicity (\Mgal $\mathrm{\simeq 10^9\,-\,10^{10}}$ \Msun, \Z\ $\mathrm{\simeq 0.1\,-\,0.3}$ \Zsun), and massive galaxies with high metallicity (\Mgal\ $\mathrm{\gtrsim 10^{11}}$\Msun, \Z\ $\mathrm{\gtrsim 0.5}$ \Zsun). The three peaks are respectively due to the high mBBH formation efficiency for low metallicities, the interplay between the SFRD and the mBBH formation efficiency for intermediate metallicities, and the high SFRD of massive galaxies. Given the current uncertainties on the global star formation history of the Universe and stellar/binary evolution, inferring the typical formation-galaxies of merging black holes should be done with caution.

Globally, the black hole mass is inversely related to metallicity due to the effect of stellar winds during the binary evolution of the progenitor stars. As a result, massive black holes (30 - 45 \Msun) tend to form in low metallicity progenitors, and small black holes (5 - 10 \Msun) arise from higher metallicity progenitors. 

When considering the progenitor formation rates of currently detectable mBBHs, we find that we can predict the progenitor environment with more confidence. The majority of detected mBBHs are from dwarf galaxies with intermediate metallicities. Across models, we see a distinct contribution from progenitors formed at high \zf\ (> 2), in  large galaxies (\Mgal $\mathrm{\gtrsim 10^{10.5}}$ \Msun) as well as a contribution from progenitors formed recently  (\zf\ $<$ 0.5) in intermediate-mass galaxies (\Mgal $\mathrm{\lesssim 10^{10}}$ \Msun). The latter arise from binaries with small delay times (< 700 Myr) while the former comes from systems with much longer delay time scales. Our progenitor posterior generation pipeline shows that GW150914 likely came from small, dwarf galaxies (\Mgal $\mathrm{\lesssim\,10^9}$ \Msun) with very low metallicity (\Z $\mathrm{\simeq 0.025}$ \Zsun).

\section*{Acknowledgments}
This work is supported by the French government, through the UCAJEDI Investments in the Future project managed by the National Research Agency (ANR) with the reference number ANR-15-IDEX-01. RS and AL acknowledge support from the graduate and research school EUR SPECTRUM. This work is also supported by the ANR COSMERGE project, grant ANR-20-CE31-001 and the "Programme National
des Hautes Energies" (PNHE) of CNRS/INSU co-funded by CEA and CNES. The authors acknowledge HPC resources from "Mesocentre SIGAMM" hosted by Observatoire de la Côte d’Azur.

\section*{Data Availability}
The data underlying this article will be shared on reasonable request to the corresponding author.

\bibliographystyle{mnras}
\bibliography{references}

\appendix
\section{BPS Flags}
\label{sec: Appendix}

\begin{table*}%[htb]
\setlength{\tabcolsep}{2pt}
\small
\begin{tabular*}{\textwidth}{p{1.75cm} p{13.5cm} p{1.8cm}}
    \toprule[1.5pt]
    \thead{\textbf{BPS Flag}}  &   \thead{\textbf{Description}}   & \thead{\textbf{Values}}\\
\midrule[1.5pt]
    \Zsun\ (zsun) & Value of solar metallicity from \citet{Asplund+2009}. & \textbf{0.014}\\
    
\midrule
    $\beta$ (beta) & Wind velocity factor $\beta_w$. -1: \citet{Belczynski+2008}; positive value: supplied to Equation 12, \citet{BSEHurley2002} . & \textbf{-1}, 0.125, 7\\
    $\dot{M}L_{\mathrm{edd}}$ \mbox{(eddlimflag)} & Wind mass-loss dependence on metallicity near the Eddington limit. 0: No dependence; 1: \citet{GiacobboMapelli2018_efficiencyBBH}. & \textbf{0}, 1\\
    
\midrule
    $\alpha$ (alpha) & CE efficiency parameter from Equation 71, \citet{BSEHurley2002}. & 0.2, 0.5, \textbf{1}, 5, 10\\
    
    $\lambda$ (lambdaf) & CE binding energy factor from Equation 69, \citet{BSEHurley2002}. 0: \citet{Claeys+2014} prescription without extra ionization energy contribution; -1: $\lambda$ chosen to satisfy $\alpha \lambda$ = 1; 1: \citet{Claeys+2014} prescription with full contribution of extra ionization energy. & -1, \textbf{0}, 1\\
    
    $CE_{\mathrm{Merger}}$ (ce\_mergeflag) & Determining whether stars beginning CE while lacking core-envelope boundary automatically lead to a stellar merger \citep{Belczynski+2008}. 0: No stellar merger; 1: stellar merger. & 0, \textbf{1}.\\
  
    $E_{\mathrm{orb,\,i}}$ \mbox{(ceflag)} & CE initial orbital energy calculation. 0: Using core mass as in Equation 70, \citet{BSEHurley2002}; 1: Using total binary mass as in \citet{deKool+1990}. & 0, \textbf{1}\\
    
    $q_{\mathrm{crit}}$ (qcflag) & CE onset based on critical mass ratio models. 0: Section 2.6 of \citet{BSEHurley2002}; 1: \citet{Hjellming_Webbink_qcrit_1and3} for GB/AGB stars, otherwise same as 0; 2: Table 2 of \citet{Claeys+2014}, 3: \citet{Hjellming_Webbink_qcrit_1and3} for GB/AGB stars otherwise same as 2; 4: Section 5.1 of \citet{Belczynski+2008}. & 0, 1, \textbf{2}, 3, 4\\

\midrule
    $v_{\mathrm{Kick}}$ \mbox{(kickflag)} & Natal kicks of Fe core-collapse supernova based on Maxwellian dispersion parameter $\sigma$ (for $v_{\mathrm{Kick}}$ = 0, -1, -2). 0: Standard COSMIC prescription with additional options from $v_{\mathrm{BH}}$; -1: Kick based on Eq. 1 of \citet{GiacobboMapelli2020}; -2: kicks based on Eq. 2 of \citet{GiacobboMapelli2020}; -3: Kicks based on Eq. 1 of \citet{BrayEldridge2016}. & \textbf{0}, -1, -2, -3\\

    $\sigma$ (sigma) & Maxwellian dispersion parameter for SN kick (km/s). & 90, \textbf{256}, 530\\
    
    $v_{BH}$ (bhflag) & Scalling BH kicks. 1: fallback-modulated \citep{Fryer+2012}; 2: Linear momentum conservation which scales down kick by BH to neutron star mass; 3: Kicks not scaled. & \textbf{1}, 2, 3\\
    
    $PISN$ (pisn) & Sets pair-instability supernova (PISN) and pulsational pair-instability supernova (PPISN) model. 0: No PISN or PPISN; -1: \citet{SperaMapelli2017}; -2: Table 1 in \citet{Marchant+2018}; -3: Table 5 in \citet{Woosley2019}. & -3,\textbf{-2}, -1, 0\\

\midrule
    $M_{\mathrm{NS-BH}}$ \mbox{(remnantflag)} & Remnant mass prescription specifically for BHs. 0: Section 6 of \citet{BSEHurley2002}; 1: \citet{Belczynski+2002}; 2: \citet{Belczynski+2008}; 3: rapid prescritopn from \citet{Fryer+2012}; 4: delayed prescription from \citet{Fryer+2012}. & 0, 1, 2, \textbf{3}, 4\\

\midrule
    $Tide_{\mathrm{ST}}$ (ST\_tide) & Tide prescription. 0: \citet{BSEHurley2002}; 1: \citet{Belczynski+2008}. & 0, \textbf{1}\\
    
\bottomrule[1.25pt]

\end{tabular*}
\caption{The description and values of the flags we explore with the COSMIC binary population synthesis (BPS) code. The three columns depict the symbol of the flag, its description, and the values we explore in our 40 simulations. Values in bold are the default values. For each BPS simulation, besides the default, we assign a given flag its value described in the table and set all other flags to their default (bold) value.}
\label{tab:BPS flag}
\end{table*}

\end{document}